\documentclass[11pt]{article}

\usepackage{tikz}
\usetikzlibrary{positioning,shapes,arrows.meta}

\usepackage[final]{acl}
\usepackage{rotating}
\usepackage{multirow}
\usepackage[normalem]{ulem}
\usepackage{graphicx}
\usepackage{subcaption}
\usepackage{float}
\usepackage{booktabs} 
\usepackage{balance}  
\usepackage{amsmath}  
\usepackage{graphicx} 

\usepackage{times}
\usepackage{latexsym}

\usepackage{amsmath} 
\usepackage{enumitem} 
\usepackage[T1]{fontenc}
\usepackage{placeins}
\usepackage{stfloats}
\usepackage[utf8]{inputenc}

\usepackage{microtype}

\usepackage{tcolorbox}
\tcbuselibrary{breakable}

\newtcolorbox{promptbox}{
  colback=gray!8,
  colframe=gray!40,
  boxrule=0.5pt,
  arc=1mm,
  breakable,
  left=3pt,
  right=3pt,
  top=3pt,
  bottom=3pt,
  fontupper=\footnotesize
}

\usepackage{inconsolata}

\usepackage{caption}

\usepackage{graphicx}
\usepackage{mdframed} 

\title{SustainableQA: A Comprehensive Question Answering Dataset for Corporate Sustainability and EU Taxonomy Reporting}

\author{
  Mohammed Ali \quad Abdelrahman Abdallah \quad Adam Jatowt \\
  University of Innsbruck, Austria \\
  \texttt{\{mohammed.ali, abdelrahman.abdallah, adam.jatowt\}@uibk.ac.at}
}

\begin{document}
\maketitle

\begin{abstract}
The growing demand for corporate sustainability transparency, particularly under new regulations like the EU Taxonomy, necessitates precise data extraction from large, unstructured corporate reports, a task for which Large Language Models and Retrieval-Augmented Generation (RAG) systems require high-quality, domain-specific question-answering datasets. To address this, we introduce SustainableQA, a novel dataset and a scalable pipeline that generates comprehensive QA pairs from corporate sustainability and annual reports by integrating semantic chunk classification, a hybrid span extraction pipeline, and a specialized table-to-paragraph transformation. To ensure high quality, the generation is followed by a novel automated assessment and refinement pipeline that systematically validates each QA pair for faithfulness and relevance, repairing or discarding low-quality entries. This results in a final, robust dataset of over 195,000 diverse factoid and non-factoid QA pairs, whose effectiveness is demonstrated by initial fine-tuning experiments where a compact 8B parameter model outperforms much larger state-of-the-art models. These results demonstrate the potential of SustainableQA as a resource for developing and benchmarking advanced knowledge assistants capable of navigating complex sustainability compliance data\footnote{\url{https://github.com/DataScienceUIBK/SustainableQA}}.
\end{abstract}

\section{Introduction}


The global financial landscape is undergoing a paradigm shift, driven by growing demands for corporate transparency in sustainability practices~\cite{venturelli2024towards,edwards2005sustainability}. Regulatory frameworks such as the European Union's Corporate Sustainability Reporting Directive (CSRD) and the EU Taxonomy for Sustainable Activities~\cite{odobavsa2023expected,hummel2024overview} require companies to provide detailed disclosures on their environmental and social impact. The CSRD mandates broad ESG reporting, while the EU Taxonomy offers a classification system for environmentally sustainable activities. These requirements have increased the length and complexity of corporate sustainability reports, which constitute a key source of information for financial and regulatory decision-making~\cite{berg2022aggregate,bronzini2024glitter,greenomy2025}. Yet such reports are often published as unstructured PDFs, making the extraction of verifiable information resource-intensive and error-prone.

The research community has increasingly turned to Large Language Models (LLMs) and Retrieval-Augmented Generation (RAG) systems 
\cite{zou2025esgreveal,wu2024susgen,bronzini2024glitter,setty2024improving,wang2024omnieval}. 
However, the efficiency of any such AI system is fundamentally dependent on the availability of high-quality, domain-specific training and evaluation data. This dependency highlights a significant research gap: the absence of publicly available, large-scale question-answering (QA) datasets specifically tailored for sustainability reporting \cite{ferjanvcivc2024textual,he2025esgenius}. This scarcity forces researchers and developers to either create small, private datasets for their experiments, a process that limits the generalizability and comparability of results or rely on general-domain models that lack the necessary expertise to navigate the nuances of regulatory terminology \cite{schimanski2024bridging}.

The creation of such a dataset is challenging due to how information is distributed within reports and the complex structure of the EU Taxonomy itself. Our analysis revealed that sustainability information is highly fragmented in corporate reports. While our initial focus was on the EU Taxonomy, we observed that crucial data points and justifications are frequently located within broader ESG and general sustainability chapters. A narrow focus on "EU Taxonomy" sections alone would therefore yield an incomplete and decontextualized set of QA pairs. This observation prompted an expansion of our scope: to build a useful resource, we must construct a comprehensive knowledge base that interconnects the three critical pillars—EU Taxonomy, ESG, and Sustainability. 

The challenge is further compounded by EU Taxonomy information being scattered across poorly formatted, multi-page tables that require extensive cross-referencing with narrative sections throughout reports. To address these difficulties, we developed a novel pipeline to generate a high-quality QA dataset from complex corporate sustainability reports. 
We employ a multi-stage process that includes: (1) semantic passage classification to identify relevant content across the entire document; (2) a hybrid span extraction pipeline that combines a fine-tuned NER model, rule-based methods, and LLM-driven refinement to ensure the precise identification of factoid answers; and (3) a specialized table-to-paragraph transformation strategy that leverages a large-context model to make complex tabular data accessible for QA generation. 

\section{Related Work}
\label{sec:related_work}

Question-answering (QA) research has evolved from general-purpose benchmarks like SQuAD \cite{rajpurkar2016squad}, Natural Questions \cite{kwiatkowski2019natural}, and HotpotQA \cite{yang2018hotpotqa} toward domain-specific datasets addressing specialized reasoning requirements.

\textbf{Corporate Finance QA.} The corporate reporting domain was pioneered by finance-focused datasets. FinQA \cite{chen2021finqa}, TAT-QA \cite{zhu2021tat}, and MultiHiertt \cite{zhao2022multihiertt} established paradigms for hybrid text-table reasoning and numerical computation over semi-structured financial documents, demonstrating the importance of multi-modal evidence integration.

\textbf{Sustainability and ESG QA.} As sustainability disclosure requirements proliferated, research adapted to ESG contexts. Systems like ChatReport \cite{ni2023chatreport} emphasized evidence provenance, while ClimRetrieve \cite{schimanski2024climretrieve} and GRI-QA \cite{contalbo2025gri} provided retrieval benchmarks and table-based QA aligned with climate disclosures and reporting standards.

\textbf{The EU Taxonomy Gap.} However, the existing work addresses voluntary ESG frameworks. The EU Taxonomy differs fundamentally by establishing legally binding disclosure requirements under the CSRD \cite{eu2022csrd}, mandating structured reporting of activity eligibility, alignment, and compliance with technical screening criteria. While conceptual frameworks for EU Taxonomy automation have been proposed \cite{hetfleisch2024automating} and limited classification datasets exist for specific sectors \cite{birti2025optimizing}, to the best of our knowledge, no public resource supports comprehensive question answering over real-world EU Taxonomy disclosures.

\textbf{SustainableQA} addresses this gap by offering more than 195,000 QA pairs derived from 61 corporate sustainability and EU Taxonomy reports, supporting both fact-based questions (e.g., ``What percentage of CapEx is aligned?'') and explanatory ones (e.g., ``Why does this activity fail DNSH?''), enabling research in hybrid reasoning, table-to-text grounding, and regulation-aware QA.


Table~\ref{tab:related_work_comparison} summarizes prior datasets; additional discussion is given in Appendix~\ref{app:related_work}.

\begin{table*}[!htbp]
\centering
\caption{Comparison of datasets, systems, and tools for sustainability reporting NLP. Scale shows dataset size with source document count in parentheses where applicable.}
\resizebox{\textwidth}{!}{%
\begin{tabular}{llllll}
\toprule
\textbf{Resource} & \textbf{Type} & \textbf{Task} & \textbf{Scale} & \textbf{Focus Area} & \textbf{Key Contribution} \\
\midrule
ChatReport & System & QA with citations & 1{,}015 reports & ESG (TCFD) & Traceable evidence provenance \\
ReportParse & Toolkit & Parsing/segmentation & --- & ESG & Unified parsing infrastructure \\
ClimRetrieve & Dataset & Retrieval benchmark & 8{,}500+ triads (30 reports) & Climate & Relevance-labeled retrieval \\
Climate Finance Bench & Dataset & QA evaluation & 330 pairs (33 reports) & Climate/ESG & Analyst-style questions in RAG \\
GRI-QA & Dataset & Table QA & 4{,}089 pairs & GRI standards & Calculated answers from GRI tables \\
ESG-Kor & Dataset & Sentence classification & 118{,}946 sentences & ESG (Korean) & Cross-lingual ESG extraction \\
CarbonPDF-QA & Dataset & PDF QA & 1{,}735 documents & Carbon footprint & Robust QA from noisy PDFs \\
ESG-Activities & Dataset & Activity classification & 1{,}325 segments & EU Taxonomy (transport) & Activity-level labeling \\
\midrule
\textbf{SustainableQA} & \textbf{Dataset} & \textbf{Multi-type QA} & \textbf{195{,}287 pairs (61 reports)} & \textbf{EU Taxonomy+ESG+Sustainability} & \textbf{Factoid, non-factoid, \& table QA} \\
\bottomrule
\end{tabular}%
}
\label{tab:related_work_comparison}
\end{table*}

\section{Dataset Construction and Analysis}
\begin{figure}[t!]
\centering
\includegraphics[width=\columnwidth]{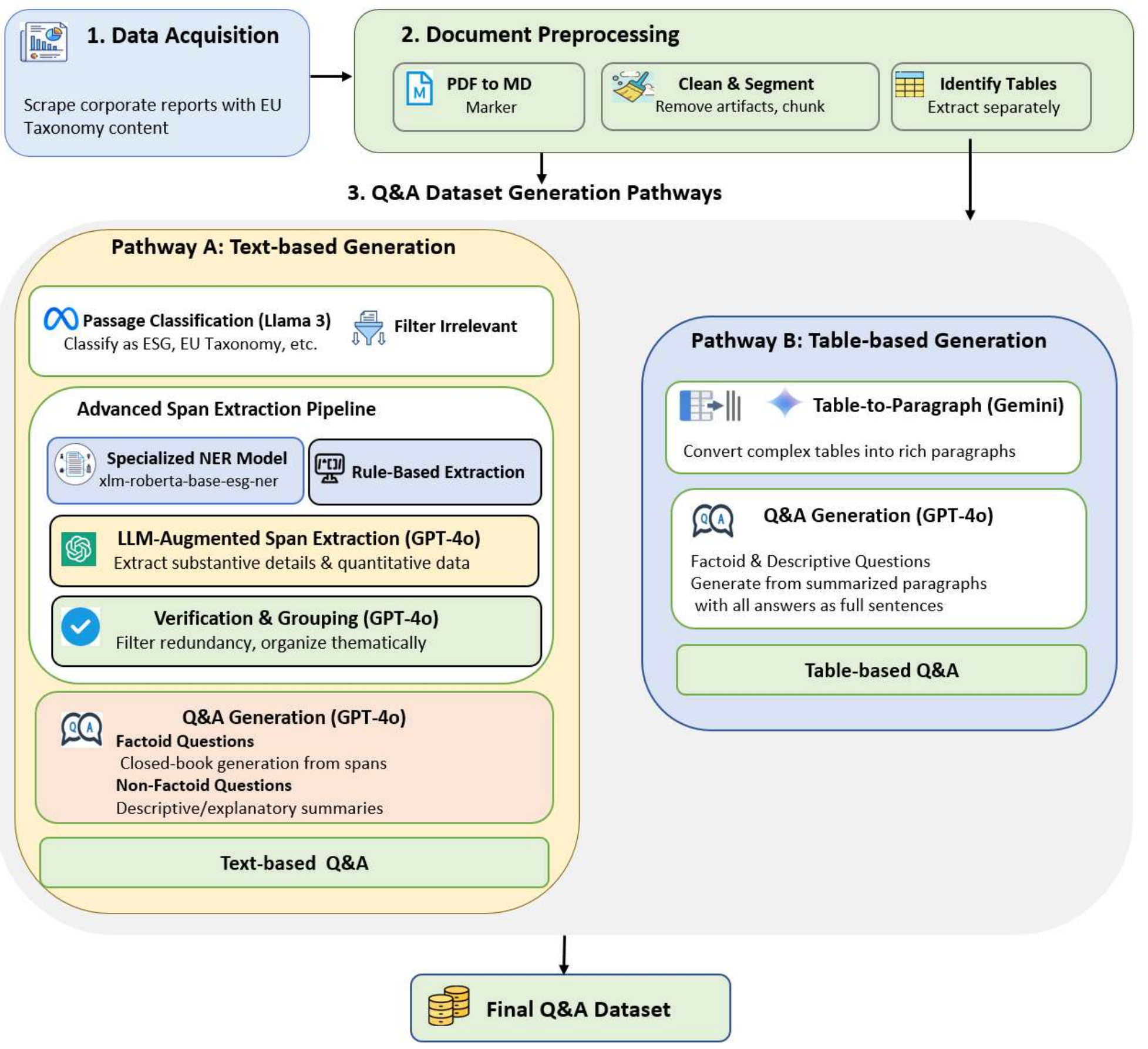}
\caption{SustainableQA dataset generation pipeline.}
\label{fig:pipeline}
\end{figure}

As shown in Figure~\ref{fig:pipeline},  the dataset pipeline is structured as a multi-stage process encompassing data acquisition, preprocessing, content classification, and question-answering generation.

\subsection{Data Acquisition}
The initial phase involves the acquisition of relevant corporate documents. This was achieved through web scraping of stock exchange websites to gather publicly available annual reports and standalone sustainability reports. We focused on companies listed on European stock exchanges with dedicated EU Taxonomy sections, prioritizing reports that also contain ESG or broader sustainability content, yielding 61 corporate reports.

\subsection{Document Preprocessing}
Raw PDF reports undergo streamlined preprocessing to convert them into a structured, clean, and manageable format. Each PDF is first transformed into Markdown text using the Marker library \cite{marker2024}, which preserves structural elements. The Markdown is then cleaned to remove non-substantive elements such as footnotes, images, and page markers, while consolidating blank lines and removing empty heading sections. Finally, the cleaned text is segmented into semantically coherent passages based on markdown headings, with a word-count constraint (e.g., \texttt{max\_words=350}) applied to ensure each passage remains within an optimal context window for subsequent LLM-based processing. In parallel, tables are identified within the Markdown output to be processed via a specialized pipeline.

\subsection{Question-Answering Dataset Generation}

The core Q\&A generation process integrates content classification, advanced span extraction, and the creation of diverse question types to maximize information coverage (see Appendix~\ref{app:prompts} for all prompts used in the generation pipeline).

\textbf{Passage Classification.}
To ensure relevance and aggregate dispersed information, each passage undergoes classification using Llama 3.3 (70B)~\cite{grattafiori2024llama} into four categories: "EU Taxonomy," "ESG," "Sustainability," or "Unknown." Unknown passages are filtered out to retain only domain-relevant content scattered throughout reports.

\subsubsection{Advanced Span Extraction Pipeline for Factoid Q\&A} 
For passages classified as relevant, a multi-stage, hybrid pipeline is used for extracting key text spans that will serve as answers for factoid questions.

\textbf{Specialized NER Model Application and Fine-tuning for Span Extraction:} The initial pass leverages a pre-trained Named Entity Recognition (NER) model, \texttt{xlm-roberta-base-esg-ner} \cite{vutukuri2023esgner}, which is designed for ESG-related entity recognition. To boost its performance, we fine-tune this model on the "ESG-only" subset of the \texttt{ExponentialScience/ESG-DLT-NER} dataset \cite{exponentialscience2023esgdlt}, focusing on \texttt{B-ESG} and \texttt{I-ESG} tags. This optimized the model's ability to accurately identify domain-specific ESG and sustainability concepts.

\textbf{Rule-Based and Dictionary-Based Span Extraction:}
To capture entities and patterns potentially missed by the NER model, we leverage regular expressions for highly structured data (e.g., regulations, standards, quantitative data). This is complemented by a spaCy \textit{PhraseMatcher} operating with comprehensive ESG and sustainability dictionaries to identify key phrases, thereby ensuring robust candidate span detection.

\textbf{Two-Stage LLM-Driven Refinement:}
The initial collected candidate spans are then subjected to a two-stage extension and refinement process using GPT 4o \cite{achiam2023gpt}. \textbf{Stage 1: LLM-Augmented Span Extraction.}
In the first stage we extend the initial set of candidate spans using LLM.
The LLM is prompted to additionally extract substantive details, quantitative data, and specific regulatory terms from the passages, explicitly excluding generic or trivial phrases. These LLM-generated spans are then aggregated with the previously collected spans (the outputs from the NER, rule-based, and \textit{PhraseMatcher} components), forming an extended set of initial candidates of answers based on which we will generate questions.

\textbf{Stage 2: Contextual Verification, Filtering, and Thematic Organization.} The candidate spans then undergo the second LLM-based processing stage that performs three functions: (1) contextual verification to ensure that the spans are actually present in the source text, (2) filtering to eliminate redundant or suboptimal entries, and (3) thematic grouping of spans into semantically coherent clusters with descriptive labels. The grouping strategy is introduced so that questions can be generated not only based on individual spans used as their answers but also on the groups of multiple spans that are semantically related. The latter helps to create complex questions which require answers composed of multiple related spans.

\subsubsection{Text-based Question-Answering}
With relevant passages identified and key spans extracted, we can now generate diverse QA pairs (see Appendix~\ref{app:qa_examples} for examples) for each passage using advanced LLMs.

\textbf{Factoid Q\&A Generation:} For every passage, we generate comprehensive factoid QA pairs using GPT-4o through a structured approach. First, we create questions based on individual spans.
Next, we create group-level questions that require multiple spans as complete answers from each thematic cluster. All questions maintain exact correspondence to their extracted passages, ensuring direct answerability from the provided context while following "closed-book" constraints, thus facilitating accurate and verifiable responses across different structural types.

\textbf{Non-Factoid Q\&A Generation:} In addition to factoid questions, we create non-factoid (descriptive/explanatory) QA pairs for each passage using GPT-4o. These questions require comprehensive textual analysis rather than isolated fact retrieval, eliciting detailed answers that explain relationships, describe processes, define concepts, or discuss implications within the passage. The generated responses typically span 1-4 sentences.

\subsubsection{Table-based Question-Answering}
Finally, a specialized approach is integrated to generate questions from tabular data.

\textbf{Table-to-Passage Transformation:} Given the large and complex tables in corporate reports, especially those related to EU Taxonomy that often span multiple pages, we convert each table into clear summarized passages using Gemini 2.5 Flash Chat \cite{google_deepmind_2025_gemini25flash}, leveraging the model's large context window capability. This process extracts essential tabular data alongside contextual information from surrounding textual content.

\textbf{QA Generation from Transformed Passages:} Following the table-to-passage transformation, we generate QA pairs with \texttt{GPT-4o}, encompassing direct numerical queries, explanatory questions about regulatory relationships, and comprehensive questions requiring integration of multiple data points. We standardize tabular answers as complete sentences to enhance evaluation robustness and preserve EU Taxonomy-specific information given the limited tabular passages relative to text passages.

\subsection{Dataset Composition and Analysis}

Our data generation and refinement pipeline produced a comprehensive dataset of 195,287 QA pairs sourced from 61 corporate reports. As detailed in Table~\ref{tab:final_stats}, the dataset is composed of 88,792 factoid (F) and 102,539 non-factoid (NF) questions derived from 8,067 text passages, complemented by 3,956 QA pairs extracted from 218 tables. The content is distributed across three key categories: ESG, EU Taxonomy, and Sustainability. A key characteristic of the dataset is the distinction in answer length: factoid answers are concise (avg. 4.2 words), while non-factoid answers are descriptive (avg. 32.4 words), targeting contextual understanding.

To assess the complexity of the generated factoid questions, we analyzed the distribution of answer spans across all 88,792 factoid QA pairs. The analysis reveals that while the vast majority (83.3\%) of questions are answered by a single, contiguous text span, the dataset also includes complex questions requiring multiple spans, with a heavy-tailed distribution extending up to 10 spans. This diversity challenges models to perform both simple entity extraction and more complex information aggregation, with over 95\% of all factoid questions answerable with four or fewer spans.

\begin{figure}[t!]
\centering
\includegraphics[width=1\columnwidth]{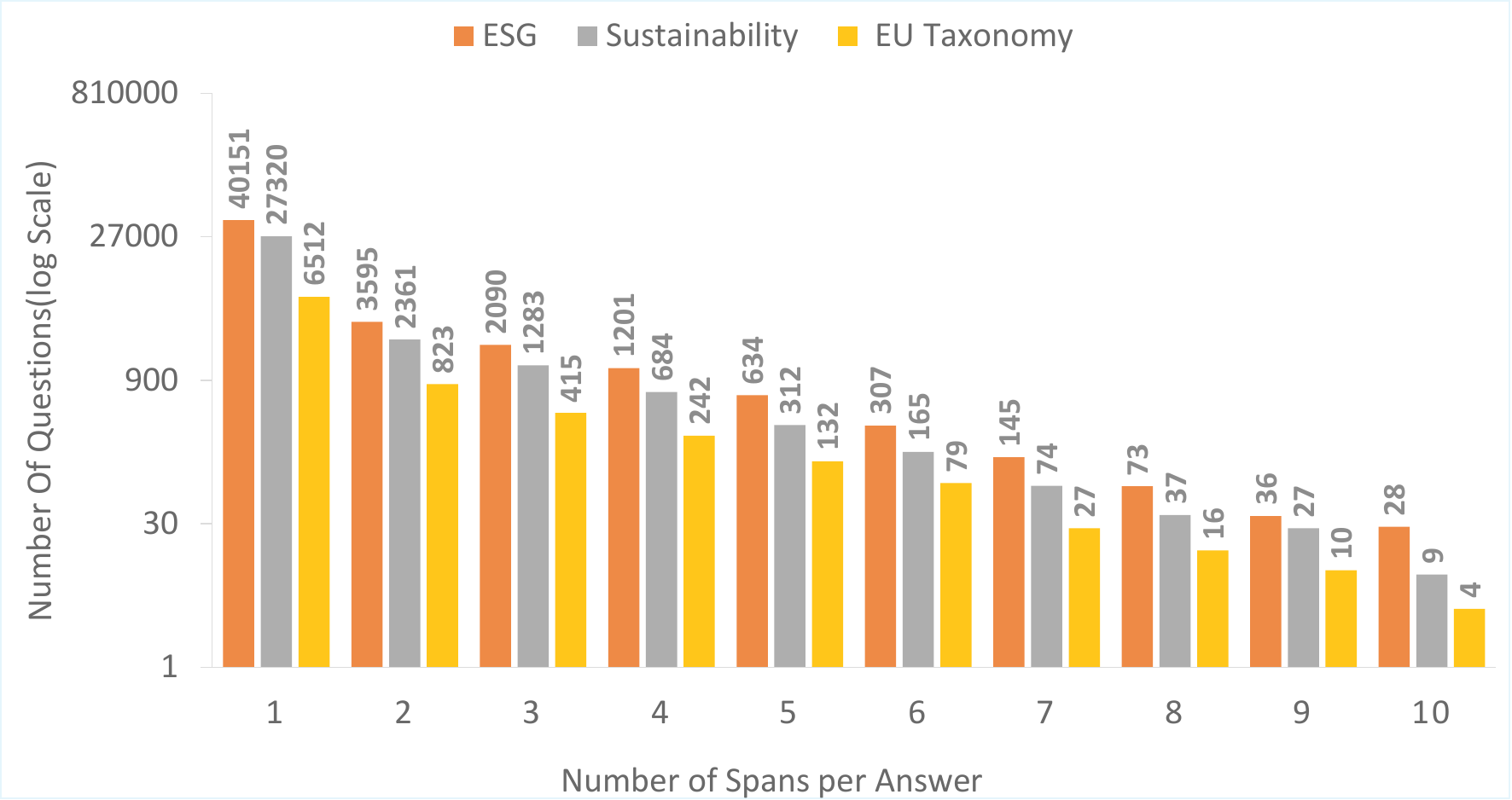}
\caption{Span distribution by category.}
\label{fig:span_by_category}
\end{figure}

When disaggregated by category (Figure~\ref{fig:span_by_category}), questions related to the EU Taxonomy are demonstrably more complex, exhibiting the highest mean number of spans (1.45) and the largest share of multi-span answers (21.2\%), as detailed in Table~\ref{tab:span_stats}. 
ESG questions show similar complexity patterns with 16.9\% requiring multiple spans, while Sustainability questions are relatively simpler with only 15.4\% multi-span answers.

\begin{table}[!htbp]
\centering
\caption{Overall dataset statistics after quality assessment and refinement (Section~\ref{sec:quality_assessment}).}
\label{tab:final_stats}
\resizebox{\columnwidth}{!}{%
\begin{tabular}{@{}lrcrcccc@{}}
\toprule
\multirow{2}{*}{\textbf{Category}} & \multirow{2}{*}{\textbf{passages}} & \multicolumn{3}{c}{\textbf{QA Pairs}} & \multicolumn{3}{c}{\textbf{Avg. Length (words)}} \\
\cmidrule(lr){3-5} \cmidrule(lr){6-8}
& & \textbf{F} & \textbf{NF} & \textbf{Total} & \textbf{Q-F} & \textbf{Q-NF} & \textbf{Ans-F/NF} \\
\midrule
ESG & 4,320 & 48,260 & 55,139 & 103,399 & 12.2 & 13.6 & 4.2/32.5 \\
EU Tax. & 747 & 8,260 & 8,906 & 17,166 & 12.7 & 14.5 & 4.7/33.5 \\
Sustain. & 3,000 & 32,272 & 38,494 & 70,766 & 12.1 & 13.4 & 4.0/32.0 \\
\midrule
\textbf{Text Subt.} & \textbf{8,067} & \textbf{88,792} & \textbf{102,539} & \textbf{191,331} & \textbf{12.2} & \textbf{13.6} & \textbf{4.2/32.4} \\
Tables & 218 & \multicolumn{2}{c}{3,956} & 3,956 & \multicolumn{2}{c}{15.8} & 23.6 \\
\midrule
\textbf{Total} & \textbf{8,285} & \multicolumn{2}{c}{---} & \textbf{195,287} & \multicolumn{3}{c}{---} \\
\bottomrule
\end{tabular}%
}
\end{table}

\begin{table}[!htbp]
\centering
\caption{Summary of answer spans for factoid questions.}
\label{tab:span_stats}
\footnotesize 
\begin{tabular}{@{}lccccc@{}}
\toprule
\textbf{Category} & \textbf{Mean} & \textbf{Med.} & \textbf{Std} & \textbf{Single} & \textbf{Multi} \\
\midrule
Overall & 1.35 & 1 & 0.97 & 83.3\% & 16.7\% \\
ESG & 1.36 & 1 & 0.99 & 83.1\% & 16.9\% \\
EU Tax. & 1.45 & 1 & 1.09 & 78.8\% & 21.2\% \\
Sustain. & 1.31 & 1 & 0.91 & 84.6\% & 15.4\% \\
\bottomrule
\end{tabular}
\end{table}

\section{Automated Quality Assessment and Refinement}
\label{sec:quality_assessment}

To ensure the reliability and utility of the dataset for training and evaluating specialized language models, we introduce a comprehensive quality assessment and refinement framework that addresses critical challenges in automatically generated QA datasets, including hallucinations, domain misalignment, and factual inconsistencies.

\begin{figure}[t]
\centering
\includegraphics[width=\columnwidth]{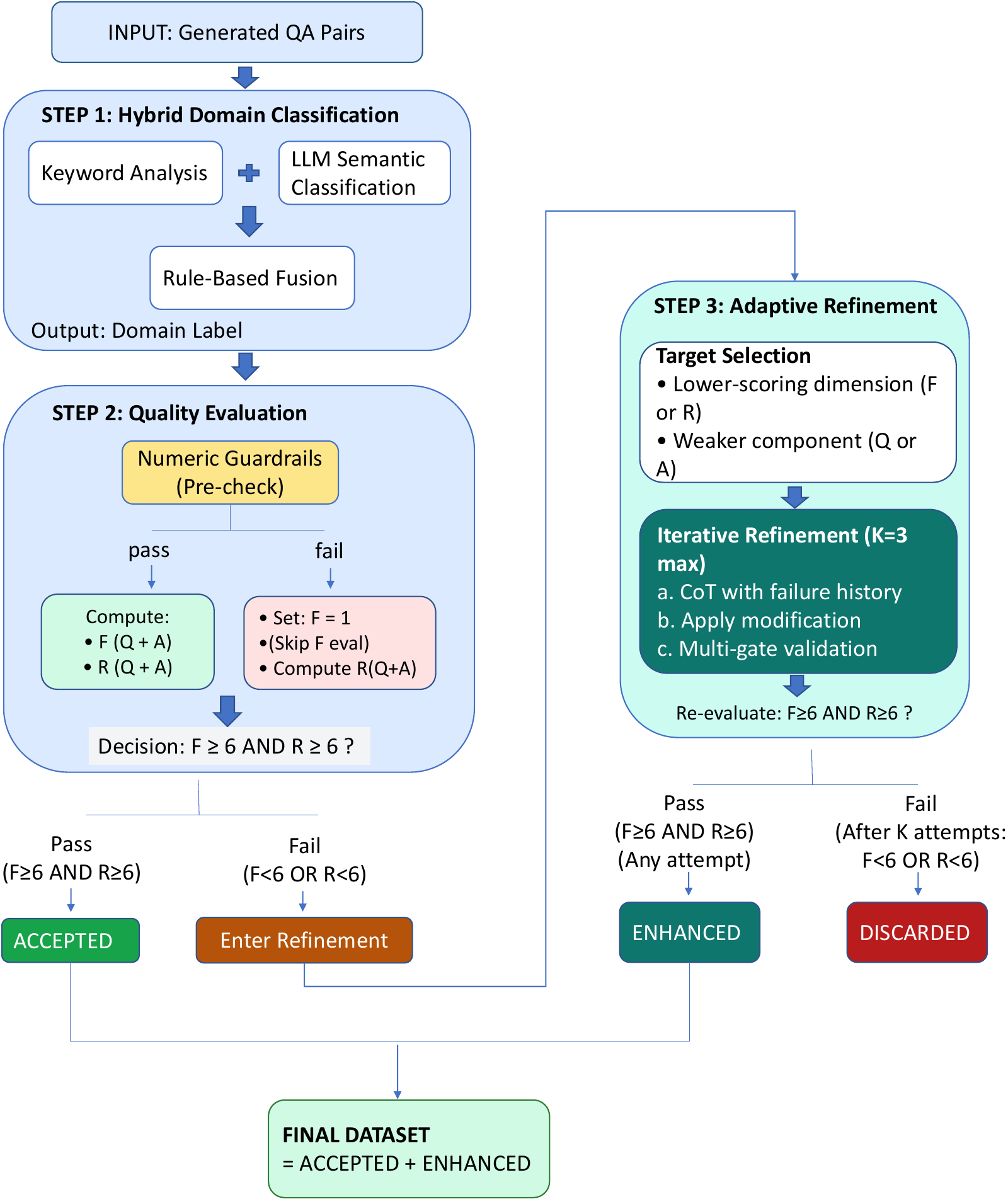}
    \caption{Three-stage quality assessment and refinement pipeline with variables: F (faithfulness score), R (relevance score), Q (question), A (answer), K (maximum refinement attempts).}
\label{fig:Assessment_pipeline}
\end{figure}

Our methodology begins with a hybrid domain classification approach that combines curated keyword analysis with LLM-based semantic understanding through rule-based fusion logic, ensuring accurate domain labeling before quality assessment, as incorrect labels would invalidate all subsequent relevance assessments (Figure~\ref{fig:Assessment_pipeline}).

We then employ a dual-metric evaluation architecture inspired by RAGAS \cite{es2023ragas} that independently assesses \textit{faithfulness} (factual grounding and freedom from hallucination) and \textit{relevance} (domain-specific utility for sustainability professionals) for both questions and answers. We use a suite of specialized LLM prompts for evaluation, modification, and correction tasks, with representative examples in Appendix~\ref{app:evaluation_prompts} and the complete set is submitted as supplementary material. The assessment adapts to question types, employing span-level analysis for factoid questions, conceptual grounding evaluation for non-factoid questions, and numerical fidelity checks for table-based questions. Both metrics use 1--10 scoring scales, with overall scores computed by averaging their respective question and answer components.

Critically, for faithfulness assessment, we implement deterministic numeric guardrails that validate quantitative claims before LLM evaluation, enforcing strict tolerances to account for rounding differences and measurement precision ($\pm$0.5 percentage points for percentages, $\leq$2\% relative error for other numerical values) to prevent hallucinations in sustainability reporting data. When numeric guardrails fail, faithfulness is set to 1.0 (faithfulness failure), and only relevance is evaluated to determine whether the pair requires faithfulness-only or dual-dimension refinement. QA pairs are accepted into the final dataset only if both overall faithfulness and relevance scores meet a configurable quality threshold (default 6.0), determined empirically to balance data quality with retention rates.

QA pairs that score below the acceptance threshold on either dimension are not discarded but instead enter an adaptive refinement pipeline designed to repair quality issues through targeted modification strategies. The system employs a hierarchical approach to identify the primary source of quality degradation, first targeting the lower-scoring dimension (faithfulness or relevance), then focusing on the weaker component (question or answer) within that dimension. For factoid answers, we implement targeted span refinement that modifies only problematic spans rather than regenerating entire answers, extracting and correcting non-verbatim or low-relevance spans individually to improve efficiency. The refinement process incorporates chain-of-thought (CoT) prompting with explicit failure awareness, inspired by self-refinement approaches \cite{madaan2023selfrefine}, where the system learns from unsuccessful attempts to construct increasingly specific prompts for subsequent retries (up to K=3 attempts).

Every modification must pass a multi-gate validation protocol (including length appropriateness, readability, coherence, and numeric consistency checks) before re-evaluation, ensuring quality improvement without introducing new errors. When all refinement attempts fail to elevate a QA pair above the acceptance threshold, the pair is deleted from the dataset, except where selective span retention applies (Appendix~\ref{sec:refinement}). 

Applying the above-described framework to 205,600 initial QA pairs achieved an overall retention rate of 95.0\%, improving dataset quality from 74.0\% initial acceptance to 89.8\% final quality across 195,287 retained pairs, with successfully refined pairs demonstrating average improvements of +1.42 for faithfulness and +1.19 for relevance. Complete methodology details, including evaluation rubrics, validation protocols, and scoring formulas, are provided in Appendix~\ref{appendix:quality_assessment}.

\section{Experiments and Results}
We conducted evaluations across multiple dimensions. First, we benchmarked retrieval methods to assess SustainableQA's utility for RAG systems. Second, we evaluated six state-of-the-art language models under three prompting strategies: zero-shot, zero-shot with context, and few-shot learning (one example per question type), and additionally fine-tuned Llama 3.1 8B on SustainableQA. The dataset was partitioned into training (80\%), validation (10\%), and test (10\%) splits, maintaining balanced category distribution.

For factoid questions, we employed span-based evaluation metrics including Exact Match (EM), Precision, Recall, and F1-score. For non-factoid and tabular questions, we utilized BERTScore, BLEU, ROUGE-L, and METEOR to evaluate semantic similarity, lexical overlap, and text quality.

Table~\ref{tab:complete_results} presents evaluation results across all models, prompting strategies, and question types, revealing several key findings:

\textbf{Fine-tuning Superiority:} The fine-tuned Llama (Llama 3.1 8B-FT, trained with context) achieved the best overall performance on 10 of the 12 metrics: 56.84\% EM, 59.70\% F1 for factoid questions, and superior performance in non-factoid (91.70\% BERTScore, 55.98\% ROUGE-L, 60.99\% METEOR) and tabular questions (75.22\% ROUGE-L, 78.76\% METEOR). The two exceptions are factoid recall, where Llama 3.3 70B with context is higher (60.76\% vs.\ 60.14\%), and tabular BERTScore, where GPT-4o with context is higher (87.24\% vs.\ 86.83\%). This demonstrates the dataset's effectiveness for domain-specific adaptation.

\textbf{Prompting Strategy Effectiveness:} Zero-shot with context consistently outperformed both pure zero-shot and few-shot approaches across all models. For factoid questions, Zero+Context achieved substantially higher exact match rates (10.12-43.15\%) compared to zero-shot (0.28-2.89\%) and few-shot (6.73-17.36\%) strategies. For non-factoid questions, Zero+Context showed dramatic improvements in BERTScore (84.61-89.97\%) versus zero-shot (74.84-81.96\%) and few-shot (81.52-86.42\%), suggesting that domain-specific context is more valuable than general examples for sustainability QA.

\textbf{Model Performance Hierarchy:} Among base models, GPT-4o and Llama 3.3 70B are the strongest in the contextual setting, although Gemma3 12B attains the highest factoid exact match (43.15\%) and Mistral 7B leads several non-factoid metrics under pure zero-shot prompting. Llama 3.3 70B achieved the highest recall (60.76\%) in factoid questions under contextual prompting, while GPT-4o led tabular question answering on BERTScore (87.24\%) and ROUGE-L (68.28\%), with Mistral 7B attaining the highest tabular METEOR (74.09\%).

\textbf{Task-Specific Challenges:} Factoid questions proved challenging, with exact match scores ranging from 0.28\% to 56.84\%, highlighting the complexity of precise span extraction in sustainability reporting. However, substantially higher F1 scores indicate that models capture relevant information despite imperfect span alignment, suggesting understanding of sustainability concepts.

\begin{table*}[htbp]
\centering
\caption{Model Performance across Question Types and Prompting Strategies. Bold = best within strategy; Underline = overall best.}
\label{tab:complete_results}
\resizebox{\textwidth}{!}{%
\begin{tabular}{@{}llcccccccccccc@{}}
\toprule
\multirow{2}{*}{\textbf{Strategy}} & \multirow{2}{*}{\textbf{Model}} & \multicolumn{4}{c}{\textbf{Factoid}} & \multicolumn{4}{c}{\textbf{Non-Factoid}} & \multicolumn{4}{c}{\textbf{Tabular}} \\
\cmidrule(lr){3-6} \cmidrule(lr){7-10} \cmidrule(lr){11-14}
& & \textbf{EM} & \textbf{F1} & \textbf{P} & \textbf{R} & \textbf{BERT} & \textbf{ROU} & \textbf{MET} & \textbf{BLU} & \textbf{BERT} & \textbf{ROU} & \textbf{MET} & \textbf{BLU} \\
\midrule
\multirow{6}{*}{Zero-shot}
& GPT-4o~\cite{achiam2023gpt} & \textbf{2.89} & \textbf{4.12} & \textbf{4.05} & 4.38 & 76.15 & 14.68 & 15.21 & 4.12 & \textbf{74.82} & \textbf{61.34} & \textbf{64.57} & \textbf{34.43} \\
& Llama 3.3 70B~\cite{grattafiori2024llama} & 1.78 & 3.64 & 3.19 & \textbf{4.71} & 78.03 & 18.93 & 21.28 & 7.89 & 73.15 & 28.59 & 37.50 & 14.40 \\
& Qwen2.5 7B~\cite{yang2025qwen3} & 0.47 & 0.62 & 0.61 & 0.65 & 79.87 & 21.98 & 24.76 & 9.49 & 71.84 & 36.11 & 45.77 & 16.59 \\
& Gemma3 12B~\cite{team2025gemma} & 2.05 & 3.48 & 3.24 & 4.10 & 81.12 & \textbf{26.19} & 27.19 & 10.48 & 72.47 & 29.49 & 31.11 & 13.95 \\
& Llama 3.1 8B~\cite{grattafiori2024llama} & 0.74 & 1.53 & 1.27 & 2.40 & 74.84 & 12.25 & 13.95 & 4.79 & 71.26 & 33.61 & 37.92 & 18.35 \\
& Mistral 7B~\cite{jiang2023mistral7b} & 0.28 & 0.54 & 0.49 & 0.68 & \textbf{81.96} & 25.28 & \textbf{29.91} & \textbf{12.12} & 71.53 & 39.95 & 47.23 & 19.14 \\
\midrule
\multirow{6}{*}{Zero+Context}
& GPT-4o~\cite{achiam2023gpt} & 41.52 & 52.19 & 50.38 & 57.94 & 89.71 & \textbf{47.90} & \textbf{56.96} & \textbf{29.29} & \underline{\textbf{87.24}} & \textbf{68.28} & 71.70 & \textbf{52.26} \\
& Llama 3.3 70B~\cite{grattafiori2024llama} & 42.89 & \textbf{54.68} & \textbf{52.75} & \underline{\textbf{60.76}} & 89.16 & 44.22 & 55.62 & 23.21 & 86.58 & 60.77 & 63.30 & 52.19 \\
& Qwen2.5 7B~\cite{yang2025qwen3} & 31.24 & 38.43 & 38.04 & 40.53 & 84.61 & 35.38 & 43.35 & 23.52 & 83.47 & 57.45 & 66.06 & 35.85 \\
& Gemma3 12B~\cite{team2025gemma} & \textbf{43.15} & 53.19 & 51.93 & 57.54 & 89.93 & 47.07 & 54.58 & 28.48 & 84.92 & 47.35 & 41.70 & 31.14 \\
& Llama 3.1 8B~\cite{grattafiori2024llama} & 32.18 & 46.80 & 43.94 & 54.85 & \textbf{89.97} & 46.10 & 54.60 & 26.22 & 84.18 & 55.25 & 55.46 & 41.83 \\
& Mistral 7B~\cite{jiang2023mistral7b} & 10.12 & 15.72 & 15.15 & 17.72 & 85.98 & 34.59 & 46.44 & 16.47 & 83.71 & 66.69 & \textbf{74.09} & 36.67 \\
\midrule
\multirow{6}{*}{Few-shot}
& GPT-4o~\cite{achiam2023gpt} & \textbf{17.36} & \textbf{23.51} & \textbf{22.74} & \textbf{24.87} & \textbf{86.42} & \textbf{34.87} & \textbf{38.94} & \textbf{18.76} & \textbf{81.29} & \textbf{64.84} & \textbf{66.71} & \textbf{37.75} \\
& Llama 3.3 70B~\cite{grattafiori2024llama} & 13.28 & 19.17 & 18.46 & 20.59 & 85.07 & 31.52 & 34.68 & 16.34 & 79.51 & 40.66 & 41.10 & 21.55 \\
& Qwen2.5 7B~\cite{yang2025qwen3} & 7.82 & 11.94 & 11.47 & 12.73 & 82.18 & 27.83 & 33.24 & 13.47 & 76.92 & 37.00 & 51.00 & 14.70 \\
& Gemma3 12B~\cite{team2025gemma} & 10.47 & 16.23 & 15.61 & 17.38 & 83.69 & 29.94 & 31.17 & 14.82 & 78.16 & 35.22 & 35.51 & 14.58 \\
& Llama 3.1 8B~\cite{grattafiori2024llama} & 8.94 & 14.76 & 14.18 & 15.83 & 81.52 & 21.74 & 23.29 & 10.18 & 76.48 & 30.52 & 29.83 & 13.87 \\
& Mistral 7B~\cite{jiang2023mistral7b} & 6.73 & 10.81 & 10.37 & 11.59 & 81.87 & 26.34 & 33.92 & 10.26 & 76.27 & 37.63 & 36.95 & 12.93 \\
\midrule
\multirow{1}{*}{Finetune}
& Llama 3.1 8B-FT & \underline{\textbf{56.84}} & \underline{\textbf{59.70}} & \underline{\textbf{59.95}} & \textbf{60.14} & \underline{\textbf{91.70}} & \underline{\textbf{55.98}} & \underline{\textbf{60.99}} & \underline{\textbf{37.99}} & \textbf{86.83} & \underline{\textbf{75.22}} & \underline{\textbf{78.76}} & \underline{\textbf{57.56}} \\
\bottomrule
\end{tabular}%
}
\end{table*}

\subsection{Retrieval Performance Benchmark}
\label{sec:retrieval_benchmark}

To validate SustainableQA's utility as a benchmark for Retrieval-Augmented Generation (RAG), we evaluated the performance of four retrieval models on the dataset's test split. We benchmarked the lexical model BM25 against three popular dense retrievers: DPR-single, BGE, and Contriever. Performance was assessed using standard metrics, including Recall@$k$, NDCG@$k$, and Mean Reciprocal Rank (MRR).

As detailed in Table~\ref{tab:retrieval_results}, the results reveal a clear performance hierarchy. The lexical retriever, BM25, achieved the strongest results across both factoid and non-factoid questions. Among the dense models, BGE was the most competitive, while DPR and Contriever showed substantially lower performance.

This performance variance highlights the distinct challenges of the sustainability domain. The success of BM25 is attributable to its strength in exact lexical matching, which is highly effective for the specialized terminology, regulatory identifiers (e.g., ``CSRD,'' ``DNSH''), and numerical KPIs prevalent in the reports. Conversely, the performance gap for dense models like DPR and Contriever may reflect two forms of mismatch:
First, a \textbf{domain mismatch}, as their general pre-training lacks exposure to specialized sustainability language. Second, a \textbf{structural mismatch}, as these models are typically trained on shorter texts ($\sim$100 words). When forced to encode our longer passages (up to 350 words) into a single fixed-size vector, the critical details required for an exact match can become less prominent, hindering precise retrieval.
BGE's stronger performance suggests that more recent dense models are better at adapting to this domain, but these results suggest that handling long-form, terminology-dense documents remains a key challenge.

\begin{table}[t]
\centering
\caption{Retrieval performance on SustainableQA test set. Best results in \textbf{bold}.}
\label{tab:retrieval_results}
\small
\setlength{\tabcolsep}{1.5pt}
\begin{tabular}{@{}llcccccc@{}}
\toprule
\textbf{Type} & \textbf{Method} & \textbf{R@1} & \textbf{R@5} & \textbf{R@10} & \textbf{N@5} & \textbf{N@10} & \textbf{MRR} \\
\midrule
\multirow{4}{*}{\small Factoid}
& BM25 & \textbf{55.4} & \textbf{75.3} & \textbf{80.6} & \textbf{66.3} & \textbf{68.0} & \textbf{64.4} \\
& BGE & 38.4 & 60.0 & 67.5 & 49.9 & 52.4 & 48.4 \\
& DPR & 12.5 & 25.1 & 31.6 & 19.0 & 21.1 & 18.9 \\
& Contriever & 12.8 & 26.3 & 33.6 & 19.8 & 22.2 & 19.7 \\
\midrule
\multirow{4}{*}{\small Non-Fact.}
& BM25 & \textbf{50.6} & \textbf{72.8} & \textbf{78.9} & \textbf{62.6} & \textbf{64.6} & \textbf{60.6} \\
& BGE & 47.4 & 70.6 & 77.2 & 59.9 & 62.0 & 57.9 \\
& DPR & 15.7 & 30.0 & 37.8 & 23.1 & 25.6 & 22.9 \\
& Contriever & 14.3 & 29.5 & 37.1 & 22.3 & 24.7 & 21.9 \\
\bottomrule
\end{tabular}
\end{table} 


\subsection{Impact of Answer Span Complexity}
To assess complexity when multi-component answers are required, we analyzed the best-performing models on factoid questions grouped by the number of answer spans (1--5 spans, covering more than 95\% of questions). As shown in Table~\ref{tab:span_degradation}, performance degrades gradually with increasing span count. Llama 3.1 8B-FT achieves 57.6\% EM on single-span questions, declining to 47.3\% on 5-span questions (17.9\% relative degradation). GPT-4o and Llama 3.3 70B exhibit similar patterns with 18.3\% and 17.5\% degradation respectively. F1 scores show greater stability with only 6--7\% degradation across all models, indicating that models maintain strong partial matching even when exact matches become harder. 

These results demonstrate that while multi-span extraction increases difficulty, the degradation remains modest and consistent across different model architectures, validating the dataset's suitability for training and evaluating QA systems on complex sustainability reporting tasks.

\begin{table*}[t!]
\centering
\caption{Performance degradation across span complexity.}
\label{tab:span_degradation}
\footnotesize
\begin{tabular}{@{}lcccccccccccc@{}}
\toprule
\multirow{2}{*}{\textbf{Model}} & \multicolumn{5}{c}{\textbf{EM (\%)}} & \multicolumn{5}{c}{\textbf{F1 (\%)}} & \multicolumn{2}{c}{\textbf{Deg 1 vs 5 (\%)}} \\
\cmidrule(lr){2-6} \cmidrule(lr){7-11} \cmidrule(lr){12-13}
& \textbf{1} & \textbf{2} & \textbf{3} & \textbf{4} & \textbf{5} & \textbf{1} & \textbf{2} & \textbf{3} & \textbf{4} & \textbf{5} & \textbf{EM} & \textbf{F1} \\
\midrule
Llama 3.1 8B-FT & 57.6 & 55.4 & 53.0 & 50.3 & 47.3 & 60.0 & 59.2 & 58.3 & 57.2 & 56.0 & 17.9 & 6.7 \\
GPT-4o & 42.1 & 40.4 & 38.6 & 36.5 & 34.4 & 52.4 & 51.8 & 51.0 & 50.1 & 49.1 & 18.3 & 6.3 \\
Llama 3.3 70B & 43.5 & 41.8 & 40.0 & 38.0 & 35.9 & 54.9 & 54.2 & 53.5 & 52.6 & 51.6 & 17.5 & 6.0 \\
\bottomrule
\end{tabular}
\end{table*}

\subsection{Human Evaluation}
To assess the dataset quality, we conducted human evaluation with three annotators who possess sustainability-related knowledge on 3,000 stratified QA pairs across ESG (1,200), EU Taxonomy (600), and Sustainability (1,200) categories, including factoid (1,800) and non-factoid (1,200) question types. 

Evaluators rated four dimensions on 5-point Likert scales, with evaluation criteria adapted to question type: \textbf{Question Quality} assessed clarity and answerability; \textbf{Answer Accuracy} evaluated exact span correctness for factoid questions and factual grounding for non-factoid questions; \textbf{Context Appropriateness} verified that answers could be derived from the provided context; and \textbf{Practical Utility} measured relevance for sustainability reporting tasks. The evaluation achieved substantial inter-annotator agreement across all dimensions: Question Quality (Krippendorff $\alpha = 0.76$), Answer Accuracy ($\alpha = 0.79$), Context Appropriateness ($\alpha = 0.74$), and Practical Utility ($\alpha = 0.71$).

Table~\ref{tab:human_eval} presents the detailed evaluation results. SustainableQA demonstrated high quality ratings with overall scores of 4.3/5.0 for Question Quality, 4.2/5.0 for Answer Accuracy, 4.1/5.0 for Context Appropriateness, and 3.9/5.0 for Practical Utility. Factoid questions consistently outperformed non-factoid variants across all dimensions (4.3 vs. 3.8 average), reflecting their greater verifiability and objective nature. Among domains, EU Taxonomy pairs achieved the highest scores across all dimensions (average 4.4/5.0), followed by ESG (4.1/5.0) and Sustainability (4.0/5.0), indicating strong alignment with regulatory reporting requirements. Answer span complexity analysis showed quality perception decline from single-span (4.5/5.0) to five-span questions (4.0/5.0), confirming that extraction difficulty correlates with human-assessed quality scores.


\begin{table}[!htbp]
\centering
\footnotesize  
\caption{Human evaluation results across dimensions, question types, and domains (5-point Likert scale).}
\label{tab:human_eval}
\setlength{\tabcolsep}{2pt}  
\renewcommand{\arraystretch}{1}
\begin{tabular}{@{}lcccc@{}}
\toprule
\textbf{Category} & \textbf{Quality} & \textbf{Accuracy} & \textbf{Context} & \textbf{Utility} \\
\midrule
\textbf{Overall} & 4.3 & 4.2 & 4.1 & 3.9 \\
\midrule
\multicolumn{5}{l}{\textit{By Question Type}}\\
\quad Factoid      & 4.5 & 4.4 & 4.3 & 4.1 \\
\quad Non-factoid  & 4.0 & 3.9 & 3.8 & 3.6 \\
\midrule
\multicolumn{5}{l}{\textit{By Domain}}\\
\quad ESG            & 4.3 & 4.2 & 4.1 & 3.9 \\
\quad EU Taxonomy    & 4.6 & 4.5 & 4.4 & 4.2 \\
\quad Sustainability & 4.2 & 4.1 & 4.0 & 3.8 \\
\bottomrule
\end{tabular}
\end{table} \vspace{-0.8em}

\subsection{Applicability}

SustainableQA is designed to support both academic and industry use cases at the intersection of AI and sustainable finance. For academic research, it provides a benchmark for training and evaluating domain-adapted language models capable of handling complex, multi-span questions grounded in EU sustainability regulations. For external stakeholders such as regulators, analysts, and investors, it enables the development of QA and RAG systems that extract and verify critical ESG and Taxonomy-aligned data with traceable evidence. Internally, companies can leverage the dataset to train AI assistants that streamline sustainability reporting and automate compliance checks under CSRD and EU Taxonomy disclosure requirements.

\section{Conclusion}
In this paper, we introduced SustainableQA, a large-scale, comprehensive QA dataset designed to address the critical need for high-quality training and evaluation data in corporate sustainability and EU Taxonomy reporting. Our comprehensive evaluations demonstrate that fine-tuning on SustainableQA enhances model performance, enabling a compact 8B parameter model to outperform much larger state-of-the-art models across different prompting strategies. The detailed analysis of answer complexity points to the challenges of multi-span extraction and establishes the dataset as a robust benchmark for evaluating model capabilities in this domain. 


\section*{Limitations}

While SustainableQA represents a significant contribution to sustainability reporting question answering, some limitations should be acknowledged. First, the dataset's scope is primarily focused on European corporate sustainability reports adhering to EU Taxonomy regulations, which may limit generalizability to other regulatory frameworks and international contexts. Second, the current dataset focuses exclusively on textual content and tabular data, excluding images, charts, and infographics that frequently appear in sustainability reports and often convey critical information such as trend visualizations and performance metrics. Future work could incorporate multimodal question answering capabilities to address visual information extraction.

\section*{Acknowledgments}
The authors would like to acknowledge the financial support provided by the Austrian Research Agency (FFG) for the project “AI Enabled Sustainability Jurisdiction Demonstrator” (project No. 915229). The computational results presented in this work have been achieved using the MUSICA cluster, part of the Austrian Scientific Computing (ASC) infrastructure.

\FloatBarrier
\bibliography{custom}
\clearpage

\appendix

\section{Detailed Related Work}
\label{app:related_work}

The evolution of question-answering (QA) systems has been fundamentally shaped by large-scale, general-purpose benchmarks that establish both task definitions and evaluation paradigms. Foundational datasets such as SQuAD \cite{rajpurkar2016squad} and Natural Questions \cite{kwiatkowski2019natural} pioneered extractive machine reading comprehension from Wikipedia articles and real search queries, respectively, while HotpotQA \cite{yang2018hotpotqa} introduced multi-hop reasoning capabilities with explicit supporting fact supervision. For retrieval-oriented QA, MS MARCO \cite{nguyen2016ms} established large-scale passage ranking and answer synthesis paradigms for open-domain settings. While these resources advanced core methodological approaches, they fail to capture the specialized linguistic patterns of regulatory discourse or the table-intensive evidence structures characteristic of corporate reporting domains.

\textbf{Finance-Specific Question Answering:}
Within corporate reporting, the finance domain emerged as the first to emphasize hybrid evidence integration (text+tables) and numerical reasoning capabilities. FinQA \cite{chen2021finqa} introduced program-supervised numerical reasoning over financial disclosures, establishing a paradigm for computational approaches to quantitative analysis. TAT-QA \cite{zhu2021tat} extended this framework to mixed tabular-textual contexts derived from authentic financial reports, while MultiHiertt \cite{zhao2022multihiertt} pioneered hierarchical multi-table reasoning combined with narrative text analysis. ConvFinQA \cite{chen2022convfinqa} introduced conversational, multi-turn chains of financial reasoning, and FinTextQA \cite{chen2024fintextqa} shifted toward long-form answers grounded in financial textbooks and regulatory sources. Collectively, these datasets established the necessity for systems capable of retrieving, computing over, and explaining information from lengthy, semi-structured documents---capabilities equally essential for sustainability reporting analysis.

\textbf{Sustainability and ESG Question Answering}

As non-financial disclosure requirements proliferated globally, research efforts adapted QA pipelines and tooling to sustainability and Environmental, Social, and Governance (ESG) reporting contexts. CHATREPORT \cite{ni2023chatreport} demonstrated LLM-based answering systems with traceable citation mechanisms over sustainability reports, emphasizing the critical importance of evidence provenance in regulatory contexts. ReportParse \cite{morio2024reportparse} provided unified parsing and segmentation infrastructure to support downstream extraction tasks, while ``Glitter or Gold?'' \cite{bronzini2024glitter} explored ESG information extraction via retrieval-augmented generation (RAG) and graph-based analytical approaches over report corpora. These contributions improved report ingestion capabilities and evidence tracing methodologies but did not provide large-scale, open QA datasets suitable for comprehensive end-to-end evaluation.

\textbf{Targeted ESG and Climate Benchmarks:}
Subsequently, more targeted benchmarks emerged to evaluate ESG-specific knowledge extraction and format handling capabilities. ClimRetrieve \cite{schimanski2024climretrieve} established retrieval benchmarks over climate-related questions grounded in corporate climate disclosures, while Climate Finance Bench \cite{mankour2025climate} targeted QA evaluation over 33 sustainability reports using analyst-style questions within RAG frameworks. GRI-QA \cite{contalbo2025gri} focused specifically on table-based QA using tables aligned with Global Reporting Initiative (GRI) standards. Complementary research explored robust QA from noisy PDF documents in carbon-footprint reporting contexts (CF-RAG/CarbonPDF-QA) \cite{zhao2025cf}. Beyond QA-focused benchmarks, ESG-Kor \cite{lee2024esg} provided 118,946 annotated sentences from Korean corporate reports for ESG sentence extraction and classification, demonstrating cross-lingual applicability of sustainability NLP methods. While these resources address retrieval evaluation, table-centric QA, and PDF robustness respectively, they fall short of providing large-scale, report-native QA datasets centered on EU-Taxonomy-relevant content.

\textbf{EU Taxonomy and Regulatory Automation:}
Unlike voluntary ESG frameworks, the EU Taxonomy establishes legally binding disclosure requirements under the CSRD. Companies are obligated to report both the eligibility and alignment of their economic activities, supported by narrative justifications, structured KPIs (e.g., turnover, CapEx, OpEx), and references to the applicable technical screening criteria (TSC) and Do-No-Significant-Harm (DNSH) requirements. While previous studies have proposed conceptual frameworks for automating EU Taxonomy compliance---often leveraging knowledge graphs or large language models (LLMs) \cite{hetfleisch2024automating}---they lack empirically grounded datasets for training or evaluation. ESG-Activities \cite{birti2025optimizing} provides 1,325 annotated text segments from non-financial disclosures labeled against EU Taxonomy activity descriptions in the transport sector, though this focuses on activity classification rather than comprehensive question answering. To the best of our knowledge, no public resource supports question answering (QA) over real-world EU Taxonomy disclosures.

\noindent\textbf{SustainableQA} fills this gap, as described in
Section~\ref{sec:related_work}. Beyond benchmarking, it enables
scalable solutions for compliance automation, verifiable reporting,
and transparent monitoring as EU sustainability regulations continue
to evolve.


\section{Automated Quality Assessment and Refinement}
\label{appendix:quality_assessment}

To ensure the reliability and utility of the dataset for training and evaluating specialized language models, we introduce a comprehensive quality assessment and refinement framework. This methodology addresses critical challenges inherent in automatically generated QA datasets, including hallucinations, domain misalignment, and factual inconsistencies, which are particularly problematic in regulatory contexts like sustainability reporting that demand high precision. The framework employs 35 specialized LLM prompts for evaluation, modification, and correction (see Appendix~\ref{app:evaluation_prompts} for representative examples; complete set is submitted as supplementary material).

\subsection{Hybrid Domain Classification}
\label{sec:domain_classification}

The initial step in our pipeline is to verify and, if necessary, correct the domain classification of each source text passage. An incorrect domain label can invalidate all subsequent relevance assessments for the QA pairs generated from that text. To address limitations in purely lexical or semantic approaches, we employ a hybrid method that combines the precision of keyword analysis with the contextual understanding of a Large Language Model (LLM).

Our method first assesses the passage against curated keyword sets for each domain. It then uses an LLM to perform a semantic classification. The final classification is determined by a rule-based fusion logic that analyzes the confidence scores from both methods. This logic is designed to be robust: \textbf{(1)} If both methods agree with high confidence, the classification is confirmed. \textbf{(2)} In cases of disagreement, the method with a higher confidence score prevails. For instance, a high-confidence LLM assessment can override a low-confidence keyword match, which is common in passages with ambiguous terminology. \textbf{(3)} If both methods exhibit low confidence, the original classification is preserved, and the item is flagged for manual expert review. This conservative approach prevents the system from making low-confidence automated judgments in the most ambiguous cases, prioritizing accuracy over automation. This hybrid approach produces a single, reliable effective classification for each passage, ensuring that all QA pairs derived from it are evaluated against a consistent and accurate domain context.

\subsection{Dual-Metric Evaluation Architecture}
\label{sec:evaluation_architecture}

Our quality assessment is built upon two orthogonal evaluation dimensions: \textit{faithfulness}, which measures the degree to which an answer is grounded in the source context, and \textit{relevance}, which assesses its alignment with the established sustainability domain. We evaluate both the question and the answer independently for each metric, and compute final scores by averaging their respective components. Both metrics are scored on a 1--10 scale. A QA pair is accepted into the final dataset only if it meets a configurable acceptance threshold (\texttt{quality\_threshold}, default 6.0), which was determined empirically to balance the need for high-quality data with reasonable dataset retention rates.

\subsubsection{Faithfulness Assessment}
\label{sec:faithfulness}

Faithfulness is a measure of factuality and freedom from hallucination. To accommodate the varied structures of generated answers, we employ type-specific evaluation strategies.

\textbf{Factoid Questions.} In extractive QA, faithfulness is synonymous with verbatim accuracy. We implement a span-level analysis where comma-separated answers are decomposed into individual spans $\{s_1, s_2, ..., s_n\}$. An LLM judge evaluates each span against the source text using a 3-tier scoring system: \textit{Score 10 (Verbatim)} for an exact match; \textit{Score 5 (Partial Match)} for spans with minor paraphrasing; and \textit{Score 1 (Non-Verbatim)} for largely paraphrased or hallucinated spans. The overall answer faithfulness is calculated using a proportional weighting formula:
\begin{equation}
    F_{\text{answer}} = \text{mean}(f_i | i \in R) \times \frac{|R|}{|T|}
\end{equation}
where $T$ is the set of all spans, $R$ is the subset of ``faithful'' spans (scoring $\ge$ 6), and $f_i$ is the faithfulness score of a span $i$. This ensures that an answer composed of many high-quality spans scores better than one with fewer, even if their average scores are similar.

\textbf{Non-factoid Questions.} For answers that require synthesis or explanation, we assess conceptual grounding. The evaluation determines whether all claims and inferences in the answer can be reasonably and logically derived from the information present in the source context, without requiring external knowledge.

\textbf{Table Questions.} For answers generated from tabular data, the evaluation prioritizes numerical and categorical fidelity. It verifies that all data points, labels, and relationships mentioned in the answer are an accurate representation of the information in the source table.

The question's faithfulness is also evaluated independently to assess whether it is answerable from the given context. The overall faithfulness score is the average of the question and answer scores:
\begin{equation}
    F_{\text{overall}} = \frac{F_{\text{question}} + F_{\text{answer}}}{2}
\end{equation}

\textbf{Deterministic Numeric Guardrails.} Given the critical importance of quantitative data in sustainability reporting, we implement deterministic pre-LLM checks, or ``guardrails,'' to prevent numerical hallucinations. For example, for an answer stating ``Taxonomy-aligned CapEx decreased by 15\% in 2023,'' our guardrail extracts the numeric claim and verifies its consistency with the source context. To do so, we apply two validation rules. First, for percentage-based claims, we allow a deviation of up to 0.5 percentage points to account for rounding differences. Second, for general numerical values such as emissions, revenue, or energy use, we require unit compatibility and enforce a maximum relative error of 2\%, calculated as the absolute difference between answer and context divided by the context value (with a small constant to avoid division by zero):

\begin{equation}
    \text{Relative Error} = \frac{|v_{\text{answer}} - v_{\text{context}}|}{\max(|v_{\text{context}}|, \epsilon)} \leq 0.02
\end{equation}
If a numeric claim fails this validation, its faithfulness score is automatically set to 1.0. This \textit{Guardrail Penalty} ensures that any QA pair with factually incorrect numbers is immediately flagged.

\subsubsection{Relevance Assessment}
\label{sec:relevance}

Relevance evaluation measures the domain-specific utility and appropriateness of a QA pair for a sustainability professional. This assessment is guided by the effective classification determined in Section~\ref{sec:domain_classification}. The core principle is to determine if a QA pair provides meaningful, domain-specific insight. A key distinction from faithfulness is that relevance is query-dependent.

\textbf{Factoid Questions.} We employ a context-aware span analysis where each span is scored based on its contribution to answering the specific question within the target domain. For example, in the ESG domain, a span like ``Scope 1 emissions'' would receive a high relevance score, whereas a generic span like ``our corporate headquarters'' would score low. The overall answer relevance is calculated using the same proportional weighting formula as faithfulness.

\textbf{Non-factoid Questions.} Relevance is assessed based on their conceptual depth and alignment with the domain's strategic priorities. A high-scoring QA pair might explore the rationale behind a company's climate strategy (Sustainability), the implementation of its governance policies (ESG), or its process for ensuring DNSH compliance (EU Taxonomy).

\textbf{Table Questions.} Relevance is determined by whether the question targets key performance indicators (KPIs) central to sustainability reporting. A question about the percentage of taxonomy-aligned revenue would be highly relevant, while a question about the number of rows in the table would not.

The final overall relevance score is the average of the question and answer relevance scores:
\begin{equation}
    R_{\text{overall}} = \frac{R_{\text{question}} + R_{\text{answer}}}{2}
\end{equation}

\subsubsection{Final Decision Rule}

The retention decision follows a strict conjunctive rule, requiring that a QA pair demonstrate high quality across both dimensions to be included in the final dataset:

\begin{multline}
    \text{RETAIN} \iff F_{\text{overall}} \ge \texttt{quality\_threshold} \\
    \land\; R_{\text{overall}} \ge \texttt{quality\_threshold}
\end{multline}

\subsection{Adaptive Refinement Pipeline}
\label{sec:refinement}

QA pairs that score below the quality threshold on either faithfulness or relevance are not immediately discarded. Instead, they enter a sophisticated refinement pipeline that attempts to repair their quality issues using chain-of-thought (CoT) learning and targeted modification strategies.

\subsubsection{Modification and Validation}

\textbf{Modification Target Selection.} To maximize efficiency, we identify the primary source of the quality issue using a hierarchical approach. The system first targets the lower-scoring dimension (faithfulness or relevance) and then, within that dimension, identifies the weaker component (question vs. answer) as the primary target for modification.

\textbf{Chain-of-Thought Refinement.} We observe that modification failures often follow repeatable patterns. Therefore, we implement a failure-aware CoT prompting strategy that explicitly learns from previous, unsuccessful refinement attempts. The system maintains a history of failures and constructs increasingly specific prompts for subsequent retries, with a bounded attempt budget (up to K attempts; default K=3).

\textbf{Targeted Span Refinement.} For factoid answers, regenerating the entire answer is inefficient. Instead, we use a targeted approach that fixes only problematic spans. The system identifies spans with low scores (non-verbatim text or weak domain relevance), then repairs each individually by correcting the text to match the source exactly or replacing it with better alternatives from context.

After each repair attempt, the modified content must first pass validation checks (detailed below) before evaluation. Once validated, we evaluate the corrected span—if the span score exceeds the threshold, we recompute the overall answer scores (faithfulness and relevance as defined in Section~\ref{sec:evaluation_architecture}). If both overall scores now meet acceptance criteria, we adopt the modified answer (KEEP\_ENHANCED). If not, we continue repair attempts up to K iterations.

When all K attempts fail to improve a span above threshold, we apply selective retention: remove the problematic span and re-evaluate the remaining answer. Due to the proportional weighting formula (Equation~1), even high-quality spans can be penalized by low-scoring ones. For example, in the relevance dimension, where spans are scored on the full 1--10 scale, an answer with two spans scoring 8.0 and 9.0 plus one scoring 3.0 yields $R_{\text{answer}} = 8.5 \times \frac{2}{3} = 5.7$ despite two strong spans; removing the problematic span raises $R_{\text{answer}}$ to 8.5, which can lift $R_{\text{overall}}$ back above the acceptance threshold. If the reduced answer achieves acceptable overall scores, we keep the QA pair with the problematic span removed. If overall scores remain below threshold even after removing bad spans, we discard the entire QA pair (DELETE\_LOW\_QUALITY).

\textbf{Validation Protocol.} To promote quality improvement while reducing the risk of introducing new errors, every modified QA pair must pass a series of automated validation checks before re-scoring. A failure at any check results in immediate rejection of the modification, counting as one failed attempt within the K-attempt budget. \textit{Gate 1 (Length Appropriateness)}—A rule-based check ensuring the modification is not unnaturally long or short. \textit{Gate 2 (Readability)}—An LLM-based check assessing linguistic quality, clarity, and grammatical correctness. \textit{Gate 3 (Coherence)}—An LLM-based check validating the logical consistency between the modified question and answer. \textit{Gate 4 (Numeric Consistency)}—A final re-application of the numeric guardrails to ensure no new quantitative errors were introduced.

\subsection{Framework Application Results}

\textbf{Implementation Setup.} We implemented our framework using Llama-3.3-70B-Instruct as the primary LLM judge with a quality threshold of 6.0 on the 1-10 scale. The initial dataset comprised 205,600 QA pairs before quality assessment.

\textbf{Initial Quality Assessment.} Before refinement, 74.0\% (152,144 pairs) met the acceptance threshold, with notable variation across categories. EU Taxonomy pairs demonstrated highest initial quality (81.2\%), followed by ESG (75.1\%) and Sustainability (70.4\%). Table-based questions achieved strongest performance (83.7\%), largely reflecting EU Taxonomy tabular data accuracy, while text-based factoid (73.5\%) and non-factoid (74.1\%) pairs showed moderate quality. Initial faithfulness scores (mean: 7.3) exceeded relevance scores (mean: 6.9) across all categories, with EU Taxonomy achieving highest scores in both dimensions (faithfulness: 7.8, relevance: 7.4).

\textbf{Refinement Outcomes.} The pipeline processed 53,456 QA pairs failing initial thresholds. Of these, 43,143 pairs (80.7\%) were successfully enhanced to meet acceptance criteria through targeted refinement, while 10,313 pairs (19.3\%) were deleted due to irreparable quality issues. The final dataset contains 195,287 QA pairs with 89.8\% overall quality, achieving 93.4\% for EU Taxonomy, 90.5\% for ESG, and 87.9\% for Sustainability. Successfully refined pairs showed average improvements of +1.42 for faithfulness and +1.19 for relevance when targeted, demonstrating the refinement pipeline's effectiveness with an overall retention rate of 95.0\%.

\section{Generation Pipeline Prompts}
\label{app:prompts}

This section provides the complete prompts used in the dataset generation pipeline.

\subsection{Classification Prompt}

\begin{promptbox}
\textbf{System:} You are an expert assistant specializing in accurately classifying text chunks related to corporate reporting into predefined categories: EU Taxonomy, ESG, Sustainability, or Unknown. Your response must be only the single chosen category label.

\medskip
\textbf{INPUT}

Title: \{title or 'N/A'\}\\
Subtitle: \{subtitle or 'N/A'\}\\
Context: \{content\}

\medskip
\textbf{TASK}

Based primarily on the Context provided, classify this text chunk into one of the following categories. Use the Title and Subtitle for additional context if needed. Return *only* the single, most appropriate class label (no explanation, reasoning, or punctuation). If specific terms (e.g., EU Taxonomy regulations) are present, prioritize them over broader themes.

\medskip
\textbf{CATEGORIES}
\begin{itemize}[nosep,leftmargin=*]
\item \textit{EU Taxonomy:} Specifically mentions EU Taxonomy regulations, principles, alignment criteria, eligibility, screening criteria, DNSH (Do No Significant Harm), Minimum Safeguards, or related KPIs (CapEx, OpEx, Turnover alignment).
\item \textit{ESG:} Focuses on Environmental, Social, or Governance factors, metrics, risks, opportunities, reporting frameworks (like GRI, SASB, TCFD, CSRD), materiality assessments, stakeholder engagement, policies, or specific ESG initiatives.
\item \textit{Sustainability:} Covers broader sustainability topics like circular economy, climate action goals, biodiversity efforts, resource efficiency, sustainable products/practices, or general corporate responsibility themes.
\item \textit{Unknown:} Does not clearly fit into any of the above categories or lacks sufficient information for classification.
\end{itemize}
\end{promptbox}

\subsection{Span Extraction Pass A}

\begin{promptbox}
\textbf{System:} You are a span extractor that prioritizes recall and outputs results in JSON format. Your goal is to find as many valid spans as possible, adhering strictly to the formatting and content guidelines.

\medskip
\textbf{TASK}

Extract concise, meaningful spans relevant to a given classification.

\medskip
\textbf{INPUT}

Context: "\{content\}"\\
Classification: "\{classification\}"

\medskip
\textbf{EXTRACTION RULES}

\textit{Include These:}
\begin{itemize}[nosep,leftmargin=*]
\item Entities, criteria, activities, and technical concepts
\item Quantitative data (e.g., '50\% reduction', '10 MWh', '€5 million')
\item Specific dates/timeframes (e.g., 'by 2030', 'FY2023')
\item Named standards/initiatives (e.g., 'GRI Standards', 'TCFD recommendations')
\item Specific regulations/policies (e.g., 'CSRD Article 8', 'EU Taxonomy')
\item Clearly defined risks or targets (e.g., 'net-zero target')
\end{itemize}

\textit{Exclude These:}
\begin{itemize}[nosep,leftmargin=*]
\item Generic terms like 'company', 'organization', 'report', 'year', 'data', 'information', 'impacts', 'approach', 'process', 'management', 'performance', 'strategy', 'framework'
\item Generic adjectives like 'important', 'significant' unless part of a named entity
\item Common functional phrases like 'in order to', 'as well as', 'responsible for'
\end{itemize}

\medskip
\textbf{FORMAT REQUIREMENTS}
\begin{itemize}[nosep,leftmargin=*]
\item Extract spans verbatim. Most spans should be 1-5 words
\item Critical named entities or specific metrics may be 6-7 words if necessary
\item Do not modify, rephrase, or summarize
\item Prioritize the most complete and specific verbatim phrase
\end{itemize}

\medskip
\textbf{OUTPUT}

\{\{ "spans": ["<span 1>", "<span 2>", ...] \}\}
\end{promptbox}

\subsection{Span Grouping Pass B}

\begin{promptbox}
\textbf{System:} You are a highly precise data structuring assistant. Your SOLE task is to critically evaluate candidate spans, filter them based on context and rules, group the valid ones, and output ONLY a valid JSON object in the specified format.

\medskip
\textbf{TASK}

Organize spans into meaningful, specific thematic groups, minimizing use of the "Individual" category.

\medskip
\textbf{INPUT}

Context: "\{content\}"\\
Classification: "\{classification\}"\\
Candidate Spans: "\{candidate\_lines\}"

\medskip
\textbf{PROCESS (SEQUENTIAL)}

\textit{Step 1: Filter \& Consolidate}
\begin{itemize}[nosep,leftmargin=*]
\item Keep only spans found verbatim in Context (case-insensitive, original casing)
\item Remove trivial/generic standalone spans
\item If spans are redundant (substrings, acronym/expansion, exact synonyms), keep only the single most complete version
\end{itemize}

\textit{Step 2: Thematic Grouping (Primary Focus)}
\begin{itemize}[nosep,leftmargin=*]
\item Actively group spans sharing a clear, specific, meaningful common theme
\item Form groups even with 2-3 highly related spans
\item Create a concise (2-5 words), highly specific 'label' for each group
\item Prioritize creating thematic groups
\end{itemize}

\textit{Step 3: "Individual" Category (Use Sparingly)}
\begin{itemize}[nosep,leftmargin=*]
\item Only truly disparate spans that cannot be thematically linked go here
\item Double-check before assigning to "Individual"
\end{itemize}

\medskip
\textbf{OUTPUT FORMAT}

Return ONLY one valid JSON object:

\{\{\\
\hspace{0.5cm}"groups": [\\
\hspace{1cm}\{\{\\
\hspace{1.5cm}"label": "<Group Label>",\\
\hspace{1.5cm}"spans": ["<Span 1>", "<Span 2>"]\\
\hspace{1cm}\}\},\\
\hspace{1cm}\{\{\\
\hspace{1.5cm}"label": "Individual",\\
\hspace{1.5cm}"spans": ["<Span>"]\\
\hspace{1cm}\}\}\\
\hspace{0.5cm}]\\
\}\}
\end{promptbox}

\subsection{Factoid Q\&A Generation}

\begin{promptbox}
\textbf{System:} You are an expert Q\&A generator. You create factoid questions strictly based on provided text and spans, ensuring the spans are the exact verbatim answers. You output only valid JSON.

\medskip
\textbf{TASK}

Generate high-quality, factoid Question-Answer pairs based strictly and solely on the provided Context.

\medskip
\textbf{INPUT DATA}
\begin{itemize}[nosep,leftmargin=*]
\item Context: "\{content\}" (The source text)
\item Spans (JSON): \{json.dumps(spans)\} (ONLY valid answers)
\item Classification: \{classification\} (Required tag)
\end{itemize}

\medskip
\textbf{CORE RULES (NON-NEGOTIABLE)}
\begin{enumerate}[nosep,leftmargin=*]
\item \textit{Answer Validity:} The 'answer' field MUST be EXACT, VERBATIM SPAN(S) ONLY from the input Spans. NO additions, omissions, paraphrasing, or changes.
\item \textit{Contextual Accuracy:} The 'answer' MUST be the PRECISE, COMPLETE, CORRECT response to the 'question', based STRICTLY on the provided Context.
\item \textit{Conditional Generation:} Generate ONLY IF you can form a natural, clear question for which the span(s) are the perfect, complete, verbatim answer. Prioritize accuracy over quantity.
\item \textit{Metadata:} Each output object MUST include "type": "Factoid" and "tag": "\{classification\}".
\end{enumerate}

\medskip
\textbf{GENERATION STRATEGY}

\textit{Span Handling:}
\begin{itemize}[nosep,leftmargin=*]
\item For each group (excluding "Individual"): create one question answered by ALL spans combined (e.g., "SpanA, SpanB"); create separate questions for EACH span individually
\item For each "Individual" span: create one question for that single span
\end{itemize}

\textit{Question Formulation:} Questions must be grammatical, natural, standalone, and vary in structure (What, When, Which, Percentage, Value, etc.).

\medskip
\textbf{EXPLICITLY AVOID}
\begin{itemize}[nosep,leftmargin=*]
\item Trivial questions (answer obvious from span alone)
\item Questions embedding the answer text
\item "What is [span]?" questions unless Context gives explicit definition
\item "Why" or "How" questions
\end{itemize}

\medskip
\textbf{VALIDATION CHECKLIST}

Before finalizing, review EVERY pair:
\begin{enumerate}[nosep,leftmargin=*]
\item Does 'answer' meet Rule 1 (EXACT VERBATIM)? (Y/N)
\item Does 'answer' meet Rule 2 (PRECISE, COMPLETE, CORRECT)? (Y/N)
\item Does 'question' meet Rule 3 and avoid all "Explicitly Avoid" types? (Y/N)
\end{enumerate}
DISCARD any pair failing any check.

\medskip
\textbf{OUTPUT}

\{\{"qa\_pairs": [...]\}\}

Each object: "question", "answer", "type": "Factoid", "tag": "\{classification\}".
\end{promptbox}

\subsection{Table Q\&A Generation}

\begin{promptbox}
\textbf{System:} You are an expert AI assistant specialized in analyzing EU Taxonomy financial reports. Your task is to generate comprehensive Question-Answering pairs based only on the provided paragraph text.

\medskip
\textbf{INTERNAL THOUGHT PROCESS}

(Do Not Show in Output)
\begin{enumerate}[nosep,leftmargin=*]
\item \textit{Understand Context:} Identify all explicit numerical data (percentages, monetary values), key terms, classifications (Taxonomy-aligned, eligible, non-eligible, transitional, enabling), specific economic activities, and environmental objectives (CCM, CCA, WTR, CE, PPC, BIO).
\item \textit{Identify Implicit Information:} Identify relationships, definitions, reasons, or processes. For example, how different figures relate or what a classification implies.
\item \textit{Categorize Information:} Systematically categorize by EU Taxonomy objective and alignment status (e.g., all turnover data, then CAPEX, then OPEX, then specific activities).
\item \textit{Formulate Diverse Questions:} Ensure a mix of: Factoid questions (specific numbers, percentages, names); Explanatory questions (relationships, purposes, implications, definitions); Holistic questions (synthesizing information from multiple parts).
\item \textit{Focus on Activities \& Statuses:} Ask specific questions about which activities are classified under each Taxonomy status (aligned, eligible, non-eligible, transitional, enabling) with their associated figures/percentages.
\item \textit{Self-Correction/Validation:} Cross-reference each Q\&A pair against the original paragraph. Verify: The question is non-trivial and specific; The answer is comprehensive, accurate, and directly supported; The Q\&A pair is distinct; Avoid questions about table title or caption.
\end{enumerate}

\medskip
\textbf{KEY GUIDELINES}
\begin{enumerate}[nosep,leftmargin=*]
\item \textit{Source Constraint:} ALL answers MUST be directly extracted or logically inferred SOLELY from the provided paragraph. Do NOT use outside knowledge.
\item \textit{Comprehensiveness:} Generate questions covering EVERY piece of information (explicit and implicit).
\item \textit{Holistic Questions:} Include questions requiring synthesis of information from multiple parts. This should be the last question, summarizing the core essence.
\item \textit{Specific Activity Focus:} Ask specific questions about Taxonomy-aligned, Taxonomy-eligible, or Taxonomy-non-eligible activities, including 'Transitional' or 'Enabling' designations with their figures/percentages.
\end{enumerate}

\medskip
\textbf{OUTPUT}

Provide a single JSON object: \{"qa\_pairs": [...]\}

Each object: "question" (string), "answer" (string). No preceding or trailing text.
\end{promptbox}

\subsection{Non-Factoid Q\&A Generation}

\begin{promptbox}
\textbf{System:} You are an expert Q\&A generator. You create descriptive, non-factoid questions and concise answers based strictly on the provided text. You output only valid JSON.

\medskip
\textbf{TASK}

Generate 7-15 high-quality, distinct, non-factoid (descriptive/explanatory) Question-Answer pairs STRICTLY from the provided Context.

\medskip
\textbf{INPUT}
\begin{itemize}[nosep,leftmargin=*]
\item Context: "\{content\}"
\item Classification: "\{classification\}"
\end{itemize}

\medskip
\textbf{CORE CONSTRAINTS}
\begin{enumerate}[nosep,leftmargin=*]
\item \textit{Strict Context Adherence:} ALL content MUST be derived SOLELY from information EXPLICITLY STATED or DIRECTLY INFERABLE within the Context. NO external knowledge.
\item \textit{Non-Factoid Focus:} Questions MUST require descriptive/explanatory answers (relationships, purposes, definitions, processes, implications). AVOID simple factoid questions answerable by a single data point.
\item \textit{Distinctness:} Each Q\&A pair must offer unique insight. Avoid redundancy.
\item \textit{Answerability:} Every question MUST be fully answerable using ONLY the Context.
\end{enumerate}

\medskip
\textbf{QUESTION FORMULATION}

Frame questions to elicit explanations/descriptions of:
\begin{itemize}[nosep,leftmargin=*]
\item Relationships (e.g., "How does X relate to Y?")
\item Processes/Mechanisms (e.g., "Explain the mechanism for X")
\item Reasons/Justifications (e.g., "What reasons are provided for Z?")
\item Implications/Consequences (e.g., "What are the implications of X?")
\item Definitions/Clarifications of concepts within the text
\end{itemize}

For Why/How questions: Ask for explanations of purpose, reasoning, methods, or processes (e.g., "Explain the rationale behind X" or "Describe the method for X"). Use company names from Context for specificity.

\medskip
\textbf{ANSWER FORMULATION}

Answers must be comprehensive yet focused (1-4 sentences, approximately 25-70 words). Ensure the answer fully addresses the question based ONLY on the text. Avoid trivial/short-phrase answers.

\medskip
\textbf{VALIDATION CHECKLIST}

Review EACH pair before output:
\begin{enumerate}[nosep,leftmargin=*]
\item Is the question genuinely non-factoid (descriptive/explanatory)? (Y/N)
\item Is the Q\&A pair distinct from others? (Y/N)
\item Is the question fully answerable ONLY from Context? (Y/N)
\item Is the answer comprehensive, accurate, and derived ONLY from Context? (Y/N)
\end{enumerate}
DISCARD any pair failing ANY check.

\medskip
\textbf{OUTPUT}

\{\{"qa\_pairs": [...]\}\}

Each object: "question", "answer", "type": "Descriptive, non-factoid", "tag": "\{classification\}".
\end{promptbox}

\section{LLM Evaluation and Refinement Prompts}
\label{app:evaluation_prompts}

This section provides representative examples of the LLM prompts used in the quality assessment and refinement framework (Section~\ref{sec:quality_assessment} and Appendix~\ref{appendix:quality_assessment}). The complete set of 35 prompts is submitted as supplementary material.

\subsection{Domain Classification}

\begin{promptbox}
\textbf{Task:} Classify text into EU Taxonomy/ESG/Sustainability/Unknown domains using keyword-guided semantic understanding.

\medskip
\textbf{System:} You are a sustainability reporting expert. Classify the following text chunk into one of the categories below.

\medskip
\textbf{Input:}
\begin{itemize}[nosep,leftmargin=*]
\item Title: \{title or 'N/A'\}
\item Subtitle: \{subtitle or 'N/A'\}
\item Context: \{passage\_text\}
\end{itemize}

\medskip
\textbf{Categories and Key Indicators:}

\textit{EU Taxonomy:}\\
Key terms: \{taxonomy\_keywords\}\\
Focus: EU Taxonomy regulations, alignment criteria, DNSH, Minimum Safeguards, CapEx/OpEx/Turnover alignment.

\textit{ESG:}\\
Key terms: \{esg\_keywords\}\\
Focus: Environmental/Social/Governance factors, reporting frameworks (GRI, SASB, TCFD, CSRD), materiality, stakeholder engagement.

\textit{Sustainability:}\\
Key terms: \{sustainability\_keywords\}\\
Focus: Circular economy, climate action, biodiversity, resource efficiency, sustainable practices.

\textit{Unknown:} Does not fit categories or lacks sufficient information.

\medskip
\textbf{Instructions:}
\begin{enumerate}[nosep,leftmargin=*]
\item Analyze context primarily, using title/subtitle for insight
\item Prioritize EU Taxonomy if specific terms present
\item Consider semantic meaning, not just keyword presence
\item Provide confidence score (1-10) and brief reasoning
\end{enumerate}

\medskip
\textbf{Output Format:}\\
Classification: [EU Taxonomy|ESG|Sustainability|Unknown]\\
Confidence: [1-10]\\
Reasoning: [Brief explanation]
\end{promptbox}

\subsection{Faithfulness Evaluation}

\subsubsection{Factoid Question Faithfulness}

\begin{promptbox}
\textbf{Task:} Determine if factoid question can be answered with verbatim spans from context.

\medskip
\textbf{System:} You are an expert evaluator for factoid span extraction.

\medskip
\textbf{Span Extraction Rules:}
\begin{enumerate}[nosep,leftmargin=*]
\item Find exact text pieces in context that answer the question
\item Spans must be word-for-word identical (no paraphrasing)
\item Multiple spans can be combined if needed, separated by commas
\item No interpretation or external knowledge allowed
\end{enumerate}

\medskip
\textbf{Evaluation Criteria (1-10):}
\begin{itemize}[nosep,leftmargin=*]
\item 10: Perfect - Clear verbatim spans directly answer the question
\item 9: Excellent - Clear verbatim spans with minimal combination needed
\item 8: Very Good - Good verbatim spans with minor combination required
\item 7: Good - Adequate verbatim spans but require careful extraction
\item 6: Acceptable - Some verbatim spans but incomplete coverage
\item 5: Weak - Limited verbatim spans with notable gaps
\item 4: Below Average - Few relevant verbatim spans
\item 3: Poor - Very limited verbatim span availability
\item 2: Very Poor - Almost no usable verbatim spans
\item 1: Failing - No verbatim spans can answer the question
\end{itemize}

\medskip
\textbf{Context:} "\{context\}"\\
\textbf{Question:} "\{question\}"

\medskip
First identify any verbatim spans that could answer the question, then evaluate answerability.

\medskip
\textbf{Format:}\\
REASONING: [Identify specific spans found and evaluate coverage]\\
SCORE: [Number from 1-10]
\end{promptbox}

\subsubsection{Non-Factoid Answer Faithfulness}

\begin{promptbox}
\textbf{Task:} Assess if non-factoid answer derives ALL information from context with no external knowledge.

\medskip
\textbf{System:} You are a meticulous evaluator of nonfactoid answer faithfulness. Verify every claim is supported by context.

\medskip
\textbf{Requirements:}
\begin{itemize}[nosep,leftmargin=*]
\item All claims must be supported by the context
\item No external knowledge or assumptions
\item Accurate representation of context information
\item Can include reasonable inferences from context
\end{itemize}

\medskip
\textbf{Evaluation Criteria (1-10):}
\begin{itemize}[nosep,leftmargin=*]
\item 10: Perfect - All information directly from context
\item 9: Excellent - Context-based with minimal interpretive language
\item 8: Very Good - Context-based with minor interpretive language
\item 7: Good - Mostly context-based with minimal reasonable inferences
\item 6: Acceptable - Mostly context-based with some reasonable inferences
\item 5: Fair - Context-based but with notable external additions
\item 4: Below Average - Some context-based content with external additions
\item 3: Poor - Limited context grounding with significant external content
\item 2: Very Poor - Minimal context grounding, mostly external content
\item 1: Failing - Significant external content or contradicts context
\end{itemize}

\medskip
\textbf{Context:} "\{context\}"\\
\textbf{Question:} "\{question\}"\\
\textbf{Answer:} "\{answer\}"

\medskip
Check each claim in the answer against the context. Identify any external knowledge or unsupported statements.

\medskip
\textbf{Format:}\\
REASONING: [Your analysis of context grounding]\\
SCORE: [Number from 1-10]
\end{promptbox}

\subsubsection{Factoid Answer Faithfulness (Span-Level JSON)}

\begin{promptbox}
\textbf{Task:} Evaluate verbatim accuracy of each factoid answer span using 3-scale scoring.

\medskip
\textbf{System:} You are evaluating verbatim accuracy of factoid answer spans\{domain\_text\}. Check if each span appears EXACTLY word-for-word in the context.

\medskip
\textbf{Context:} \{context\}\\
\textbf{Question:} \{question\}

\medskip
\textbf{Spans to Evaluate:}
\begin{itemize}[nosep,leftmargin=*]
\item \{span\_1\}
\item \{span\_2\}
\item ...
\end{itemize}

\medskip
\textbf{Instructions:}
\begin{enumerate}[nosep,leftmargin=*]
\item For each span, check word-by-word if it appears exactly in context
\item Score using 3-scale scoring:
  \begin{itemize}[nosep]
  \item 10: PERFECT VERBATIM - all words appear exactly in context
  \item 5: PARTIAL VERBATIM - some words verbatim, some not
  \item 1: NOT VERBATIM - most/all words not found exactly in context
  \end{itemize}
\item For spans scoring < 10, identify specific verbatim problems
\end{enumerate}

\medskip
\textbf{Return strict JSON:}

\{\\
\hspace{0.5cm}"spans": [\\
\hspace{1cm}\{\\
\hspace{1.5cm}"span": "<exact span text>",\\
\hspace{1.5cm}"verbatim\_score": <10, 5, or 1>,\\
\hspace{1.5cm}"verbatim\_problem": "<issue if score < 10>",\\
\hspace{1.5cm}"non\_verbatim\_words": ["<word1>", "<word2>"],\\
\hspace{1.5cm}"found\_in\_context": true|false\\
\hspace{1cm}\}\\
\hspace{0.5cm}],\\
\hspace{0.5cm}"reasoning": "<brief verbatim assessment summary>"\\
\}

\medskip
Focus on exact word matching - number formats, terminology, punctuation must match exactly.
\end{promptbox}

\subsection{Relevance Evaluation}

\subsubsection{Factoid Answer Relevance (Span-Level JSON)}

\begin{promptbox}
\textbf{Task:} Assess domain relevance of each factoid answer span with evidence.

\medskip
\textbf{System:} You are a careful, honest expert. Return STRICT JSON only (no Markdown).

\medskip
\textbf{User Prompt:} You are an expert evaluator for factoid answer relevance in the \{classification\} domain. Your task is to assess how relevant each span is for \{classification\} analysis, using the full CONTEXT. Do NOT calculate any overall/final score. Only score each span and provide supporting evidence.

\medskip
\textbf{Context:} \{context\}\\
\textbf{Question:} \{question\}

\medskip
\textbf{Spans to Evaluate:}
\begin{itemize}[nosep,leftmargin=*]
\item \{span\_1\}
\item \{span\_2\}
\item ...
\end{itemize}

\medskip
\textbf{Domain Anchors (\{classification\}):} \{relevant\_keywords\}

\medskip
\textbf{Instructions:}
\begin{enumerate}[nosep,leftmargin=*]
\item For each span, evaluate its relevance to \{classification\} analysis based on domain value
\item Score each span 1-10:
  \begin{itemize}[nosep]
  \item 8-10: Essential \{classification\} KPI/metric with clear domain connection
  \item 6-7: Important \{classification\} data with domain relevance
  \item 4-5: General business data with weak \{classification\} connection
  \item 1-3: Administrative/operational data with no \{classification\} relevance
  \end{itemize}
\item Identify domain anchors (keywords) that support the relevance assessment
\end{enumerate}

\medskip
\textbf{Return strict JSON:}

\{\\
\hspace{0.5cm}"spans": [\\
\hspace{1cm}\{\\
\hspace{1.5cm}"span": "<span text>",\\
\hspace{1.5cm}"evidence": "<context evidence supporting \{classification\} relevance>",\\
\hspace{1.5cm}"anchors": ["relevant\_keyword1", "relevant\_keyword2"],\\
\hspace{1.5cm}"score": <1-10>\\
\hspace{1cm}\}\\
\hspace{0.5cm}],\\
\hspace{0.5cm}"reasoning": "<2-3 sentence justification based on \{classification\} domain value>"\\
\}

\medskip
Do NOT calculate or return any overall/final score. Score and explain each span individually.
\end{promptbox}

\subsubsection{Non-Factoid Answer Relevance}

\begin{promptbox}
\textbf{Task:} Evaluate answer's domain relevance prioritizing contextual accuracy and practical utility.

\medskip
\textbf{System:} You are an expert in \{classification\} strategy evaluation.

\medskip
\textbf{Classification Context:}
\begin{itemize}[nosep,leftmargin=*]
\item ESG: Environmental, Social, and Governance factors in business operations and reporting
\item EU Taxonomy: European Union's classification system for sustainable economic activities
\item Sustainability: Broader sustainable development, environmental stewardship, and corporate responsibility
\end{itemize}

\medskip
\textbf{Important:} Non-factoid answers often explain business processes, strategies, or organizational aspects that support \{classification\} goals. These can be highly relevant even without deep theoretical analysis.

\medskip
\textbf{Relevance Criteria (Prioritizing Practical Utility):}
\begin{itemize}[nosep,leftmargin=*]
\item 10: Directly answers question with clear \{classification\} context relevance and accuracy
\item 9: Well-grounded answer with strong \{classification\} connection and practical utility
\item 8: Good contextual answer with clear \{classification\} relevance for domain professionals
\item 7: Adequate contextual answer with noticeable \{classification\} connection
\item 6: Some contextual relevance with \{classification\} domain applicability
\item 5: Limited \{classification\} connection but contextually accurate
\item 4: Weak \{classification\} relevance, mostly generic but accurate content
\item 3: Minimal \{classification\} connection
\item 2: Very limited \{classification\} relevance
\item 1: Not relevant to \{classification\} domain or contextually inaccurate
\end{itemize}

\medskip
\textbf{Evaluation Principle:} Contextually accurate answers that explain business processes, strategies, or organizational aspects supporting \{classification\} goals should score well (7-9) even without deep theoretical analysis, as practical implementation is crucial for \{classification\} success.

\medskip
\textbf{Context:} "\{context\}"\\
\textbf{Question:} "\{question\}"\\
\textbf{Answer:} "\{answer\}"

\medskip
\textbf{Format:}\\
REASONING: [Explain practical \{classification\} domain utility and contextual accuracy]\\
SCORE: [Number from 1-10]
\end{promptbox}

\subsection{Question Modification}

\subsubsection{Factoid Question Faithfulness Modification}

\begin{promptbox}
\textbf{Task:} Improve factoid question to better target verbatim spans from context while preserving answer consistency.

\medskip
\textbf{System:} You are an expert in \{classification\} reporting and factoid question improvement.

\medskip
\textbf{Question Analysis:} The question "\{question\}" may have issues with verbatim answerability. Analyze what makes it difficult to answer with exact spans from context.

\medskip
\textbf{Modification Approach:}
\begin{enumerate}[nosep,leftmargin=*]
\item IDENTIFY PROBLEMS: Determine why current question is hard to answer with exact spans
\item TARGET SPECIFIC DATA: Focus on data points (numbers, names, dates) actually present in context
\item ALIGN LANGUAGE: Use terminology that matches context exactly
\item ENSURE EXTRACTABILITY: Make sure answers can be word-for-word from context
\end{enumerate}

\medskip
\textbf{Answer Consistency Constraint:} The modified question MUST target the same data as the answer: "\{answer\}". Your modified question should elicit this exact same factoid answer from the context. Ensure the question targets the specific spans/data points that produce this answer.

\medskip
\textbf{Preserve \{classification\} Relevance:}
\begin{itemize}[nosep,leftmargin=*]
\item Maintain \{classification\} domain focus where possible
\item Keep domain-specific terminology when supported by context
\item Ensure question remains valuable for \{classification\} analysis
\end{itemize}

\medskip
\textbf{Requirements:}
\begin{itemize}[nosep,leftmargin=*]
\item Question must be DIFFERENT from "\{question\}"
\item Must target verbatim extractable information
\item Should allow multiple potential spans for comprehensive answers
\item MUST elicit this specific answer: "\{answer\}"
\end{itemize}

\medskip
\textbf{Context:} "\{context\}"\\
\textbf{Current Question:} "\{question\}"\\
\textbf{Target Answer:} "\{answer\}"

\medskip
\textbf{Format:}\\
REASONING: [Analyze current question's problems, explain specific changes and why they improve verbatim answerability and preserve answer consistency]\\
MODIFIED\_QUESTION: [Substantially different question targeting verbatim extractable data that produces the target answer]
\end{promptbox}

\subsubsection{Non-Factoid Question Relevance Modification}

\begin{promptbox}
\textbf{Task:} Enhance non-factoid question's domain relevance and analytical depth while maintaining context grounding.

\medskip
\textbf{System:} You are an expert in non-factoid question modification for \{classification\} domain.

\medskip
\textbf{\{classification\} Key Terms:} \{relevant\_keywords\}

\medskip
\textbf{Question Analysis:} The question "\{question\}" may lack sufficient \{classification\} domain focus or analytical depth. Analyze what domain-specific improvements are needed.

\medskip
\textbf{Modification Approach:}
\begin{enumerate}[nosep,leftmargin=*]
\item IDENTIFY DOMAIN GAPS: Determine what \{classification\}-specific improvements are needed
\item INCORPORATE FRAMEWORKS: Use \{classification\} analytical frameworks and concepts
\item TARGET PROFESSIONAL INSIGHTS: Focus on analysis valuable for \{classification\} professionals
\item MAINTAIN ANALYTICAL NATURE: Keep explanatory/analytical character
\end{enumerate}

\medskip
\textbf{Preserve Faithfulness Quality:}
\begin{itemize}[nosep,leftmargin=*]
\item Ensure modified question can be answered from available context
\item Do not create questions requiring extensive external knowledge
\item Keep analytical nature while ensuring context provides foundation
\end{itemize}

\medskip
\textbf{Requirements:}
\begin{itemize}[nosep,leftmargin=*]
\item Question must be DIFFERENT from "\{question\}"
\item Must enhance \{classification\} professional relevance
\item Should target domain-specific analytical insights
\end{itemize}

\medskip
\textbf{Context:} "\{context\}"\\
\textbf{Current Question:} "\{question\}"

\medskip
\textbf{Format:}\\
REASONING: [Analyze current question's domain relevance, explain specific changes and why they enhance \{classification\} analytical focus]\\
MODIFIED\_QUESTION: [Substantially different question with enhanced \{classification\} domain relevance and analytical depth]
\end{promptbox}

\subsection{Answer Modification}

\subsubsection{Non-Factoid Answer Relevance Enhancement (CoT)}

\begin{promptbox}
\textbf{Task:} Enhance non-factoid answer's domain relevance while maintaining strict context faithfulness.

\medskip
\textbf{System:} You are an expert in corporate sustainability reporting and analytical answer improvement.

\medskip
\textbf{CoT Instructions:} \{cot\_instructions\}

\medskip
\textbf{Context:} "\{context\}"\\
\textbf{Question:} "\{question\}"\\
\textbf{Original Answer:} "\{answer\}"

\medskip
\textbf{\{classification\} Domain Focus:}
\begin{itemize}[nosep,leftmargin=*]
\item ESG: Focus on environmental impact, social responsibility, governance practices, stakeholder value, risk management
\item EU Taxonomy: Focus on sustainable economic activities, taxonomy alignment, technical screening criteria, environmental objectives
\item Sustainability: Focus on sustainable development, resource efficiency, long-term value creation, impact measurement
\end{itemize}

\medskip
\textbf{Enhancement Strategy:}
\begin{enumerate}[nosep,leftmargin=*]
\item ANALYTICAL DEPTH: Provide deeper analysis relevant to \{classification\} professionals and decision-makers
\item DOMAIN PERSPECTIVE: Frame insights from \{classification\} operational and strategic viewpoints
\item PROFESSIONAL VALUE: Target content that \{classification\} practitioners would find actionable
\item CONTEXTUAL GROUNDING: Keep all enhancements strictly based on provided context - do not add external information
\item FLEXIBLE APPROACH: Enhance analytical, numerical, or explanatory content as appropriate
\item CONTEXT LIMITATION: All \{classification\} domain enhancements must be derivable from the given context
\end{enumerate}

\medskip
\textbf{Modification Requirements:}
\begin{itemize}[nosep,leftmargin=*]
\item Enhance \{classification\} domain analytical perspective using only context-supported information
\item Strengthen professional value for \{classification\} stakeholders and practitioners
\item Improve strategic relevance to \{classification\} business objectives and frameworks
\item Maintain answer type (analytical/numerical/explanatory) while enhancing domain focus
\item All modifications must stay within the boundaries of the provided context
\item Answer must be DIFFERENT from original with clear \{classification\} relevance improvement
\end{itemize}

\medskip
\textbf{Format:}\\
REASONING: [Explain how \{classification\} domain relevance was enhanced using only context-supported information while preserving faithfulness]\\
MODIFIED\_ANSWER: [Enhanced answer with stronger \{classification\} domain focus derived strictly from context information]
\end{promptbox}

\subsubsection{Table Answer Faithfulness Enhancement (CoT)}

\begin{promptbox}
\textbf{Task:} Modify table answer to be perfectly faithful to table context while handling both specific and summary questions.

\medskip
\textbf{System:} You are an expert in corporate sustainability reporting and tabular data analysis.

\medskip
\textbf{CoT Instructions:} \{cot\_instructions\}

\medskip
\textbf{Table Context:} "\{context\}"\\
\{table\_caption\_info\}\\
\textbf{Question:} "\{question\}"\\
\textbf{Original Answer:} "\{answer\}"

\medskip
\textbf{Important:} The TABLE CONTEXT has been converted to paragraph format. Original table structure (rows, columns, cell positions) is not preserved in this format.

\medskip
\textbf{Table Faithfulness Requirements:}
\begin{enumerate}[nosep,leftmargin=*]
\item DATA ACCURACY: All numerical values, percentages, names, and categories must match the TABLE CONTEXT exactly
\item CALCULATION PRECISION: Any mathematical operations, totals, or derived values must be mathematically correct
\item NO INVENTED DATA: Do not create, assume, or extrapolate data points not present in the TABLE CONTEXT
\item SUMMARY ACCURACY: For summary questions, ensure aggregations and generalizations are based on actual TABLE CONTEXT
\item REFERENCE PRECISION: Accurately describe data relationships and patterns present in the TABLE CONTEXT
\end{enumerate}

\medskip
\textbf{Table Question Types Handling:}
\begin{itemize}[nosep,leftmargin=*]
\item SPECIFIC DATA EXTRACTION: Direct numerical values, names, categories from TABLE CONTEXT
\item SUMMARY/ANALYTICAL: Trends, patterns, comparisons, totals derived from TABLE CONTEXT
\item COMPARATIVE: Relationships between different table elements
\item CATEGORICAL: Groupings and classifications present in TABLE CONTEXT
\end{itemize}

\medskip
\textbf{Modification Process:}
\begin{enumerate}[nosep,leftmargin=*]
\item Cross-verify all numerical values and data points against TABLE CONTEXT
\item Correct any calculation errors or mathematical inaccuracies
\item Remove claims not supported by TABLE CONTEXT
\item For summary questions, ensure generalizations are data-based
\item Ensure answer must be DIFFERENT from original while improving data accuracy
\end{enumerate}

\medskip
\textbf{Validation Checklist:}
\begin{itemize}[nosep,leftmargin=*]
\item Do all numbers exactly match the TABLE CONTEXT?
\item Are calculations mathematically correct based on available data?
\item Are summary statements supported by actual table content?
\item Is the data interpretation accurate and complete?
\end{itemize}

\medskip
\textbf{Format:}\\
REASONING: [Explain what data accuracy issues were corrected and how the answer now faithfully represents table information]\\
MODIFIED\_ANSWER: [Accurate answer based precisely on TABLE CONTEXT, appropriate for question type]
\end{promptbox}

\subsection{Span-Level Correction}

\begin{promptbox}
\textbf{Task:} Fix non-verbatim words in factoid span using exact context text.

\medskip
\textbf{System:} Fix verbatim issues in \{classification\} spans without redundancy.

\medskip
\textbf{Context:} \{context\}\\
\textbf{Current Span:} "\{span\_text\}"\\
\textbf{Verbatim Problem:} \{verbatim\_problem\}\\
\textbf{Other Existing Spans:} \{existing\_content\}\\
\textbf{Domain:} \{classification\}

\medskip
\textbf{Requirements:}
\begin{enumerate}[nosep,leftmargin=*]
\item Fix non-verbatim words using EXACT word-for-word replacements from context
\item Keep verbatim words unchanged
\item CRITICAL: Modified span must NOT overlap with existing spans: \{existing\_content\}
\item Avoid any redundancy (maximum 40\% word overlap with existing spans)
\item Return "REMOVE" if cannot fix without creating redundancy
\end{enumerate}

\medskip
\textbf{Examples:}
\begin{itemize}[nosep,leftmargin=*]
\item Fix "2.8M" → "2,800,000" (if found exactly in context)
\item Fix "admin costs" → "administrative expenses" (if found exactly)
\item Keep verbatim words unchanged
\end{itemize}

\medskip
Return only the corrected span or "REMOVE":
\end{promptbox}



\clearpage
\onecolumn
\section{Question-Answer Examples}
\label{app:qa_examples}

\begin{table}[H]
\centering
\caption{Sample factoid (F), non-factoid (NF), and table-based (T) QA pairs from SustainableQA.}
\label{tab:qa_examples}
\footnotesize
\setlength{\tabcolsep}{4pt}
\begin{tabular}{@{}l p{5.2cm} p{9.3cm}@{}}
\toprule
\textbf{Type} & \textbf{Question} & \textbf{Answer} \\
\midrule
F & Which EU directive establishes the framework for nearly zero-energy buildings in national plans? & Directive 2010/31/EU \\
\addlinespace
F & Which types of fuels are considered renewable under the Climate Delegated Act screening criteria? & biofuels, biogas, hydrogen-based synthetic fuels \\
\addlinespace
NF & What role do DNSH criteria and minimum safeguards play in determining Taxonomy alignment? & DNSH criteria and minimum safeguards are essential for determining Taxonomy alignment as they ensure that the activity does not cause significant harm and adheres to ethical and environmental standards. However, these are only assessed if the activity meets the initial contribution criteria for climate protection. \\
\addlinespace
NF & Why does the Group focus on both CO$_2$ and non-CO$_2$ emissions in its climate strategy? & The Group focuses on both because addressing only CO$_2$ does not fully capture the overall climate impact of air traffic. Non-CO$_2$ factors, such as nitrogen oxides and the formation of condensation trails, also significantly contribute to climate change and require mitigation. \\
\addlinespace
T & Why was the Group's Taxonomy alignment reported as 0\% for 2023 despite having Taxonomy-eligible activities? & The Group was unable to demonstrate substantial contribution (SC) for Climate Change Mitigation (CCM 7.1) or Circular Economy (CE 3.1) and did not meet the ``Do No Significant Harm'' (DNSH) criteria for any of the six environmental objectives. \\
\addlinespace
T & Summarize the Group's overall EU Taxonomy financial performance for 2023. & In 2023, the Group reported no taxonomy-aligned Revenue, CapEx, or OpEx. While 58\% of its CapEx (17,807 TEUR) and 9.8\% of its OpEx (898 TEUR) were taxonomy-eligible, the majority of OpEx (90.2\%) and a significant portion of CapEx (42\%) were not taxonomy-eligible. Overall, no activities were classified as taxonomy-aligned. \\
\bottomrule
\end{tabular}
\end{table}

\end{document}